\documentclass[reprint,aps,amsmath,amssymb,floatfix,prb,superscriptaddress]{revtex4-2}
\usepackage{graphicx}% Include figure files
\usepackage{dcolumn}% Align table columns on decimal point
\usepackage{bm}% bold math
\usepackage{xr-hyper}
\usepackage{braket}
\usepackage{hyperref}

\hypersetup{
  bookmarks		= false,		% show bookmarks bar %
  unicode			= false,	% non-Latin characters in Acrobat's bookmarks
  pdfborder		= {0 0 0}	% style of the border around a link; {0 0 0} gives no border %
  pdftoolbar		= false,		% show Acrobat's toolbar %
  pdfmenubar		= true,		% show Acrobat's menu %
  pdffitwindow		= false,     	% window fit to page when opened %
  pdfstartview		= {FitH},    	% fits the width of the page to the window %
  pdfnewwindow	= true,      	% links in new window %
  colorlinks		= true,		% false: boxed links; true: colored links %
  linkcolor			= blue,		% color of internal links %
  citecolor			= blue,		% color of links to bibliography %
  filecolor			= blue,		% color of file links %
  urlcolor			= blue,		% color of external links %
  linktocpage		= true,		% whole entry or only page in the toc is made into a link %
  pdftitle			= {},
  pdfauthor		= {},
  pdfsubject		= {},
  pdfcreator		= {MikTeX},
  pdfproducer		= {},
  pdfkeywords		= {},
}

\newcommand{\vect}[1]{\boldsymbol{#1}}		% I use command \vect{} for vectors
\newcommand{\op}[1]{\hat{\boldsymbol{#1}}}	% I use command \op{} for operators

\newcommand{\epsilonLT}{\epsilon_{\mathrm{LT}}}
\newcommand{\qvector}{\vect{q}_{\mathrm{CDW}}}

\newcommand{\HTB}{\op{H}_{\mathrm{TB}}}
\newcommand{\HD}{\op{H}}
\newcommand{\HCDW}{\op{H}_{(2\times2)}}
\newcommand{\hLT}{\op{h}}
\newcommand{\hLTCDW}{\op{h}_{(2\times2)}}
\newcommand{\threebythree}{(3\times3)}
\newcommand{\twobytwo}{(2\times2)}
\newcommand{\sqrtthree}{(2\sqrt{3}\times2\sqrt{3})R30^{\circ}}
\newcommand{\gapmatrix}{\op{\Delta}}
\newcommand{\gap}{\tilde{\Delta}}
\newcommand{\gaptwo}{\Delta}
\newcommand{\added}[1]{\textcolor{black}{#1}}

\begin{document}

\preprint{APS/123-QED}

\title{Controlling charge density order in 2H-TaSe$_{2}$ using a van Hove singularity}
\author{W.~R.~B.~Luckin} % wl748@bath.ac.uk; will@luckin.co.uk
\thanks{W.~R.~B.~L.~and Y.~L.~contributed equally to this work.}
%\altaffiliation{These authors contributed equally to this work.}
\affiliation{Department of Physics, University of Bath, Claverton Down, Bath BA2 7AY, United Kingdom}
\author{Y.~Li} % liyw3@shanghaitech.edu.cn
\thanks{W.~R.~B.~L.~and Y.~L.~contributed equally to this work.}
\email{yiweili@whu.edu.cn}
%\altaffiliation{These authors contributed equally to this work.}
\affiliation{Institute for Advanced Studies (IAS), Wuhan University, Wuhan 430072, People’s Republic of China}
\affiliation{School of Physical Science and Technology, ShanghaiTech University, Shanghai 201210, People’s Republic of China}
\author{J.~Jiang} % jjiangcindy@ustc.edu.cn
% \affiliation{School of Physical Science and Technology, ShanghaiTech University, Shanghai 201210, People’s Republic of China}
% \affiliation{CAS-Shanghai Science Research Center, Shanghai 201210, People’s Republic of China}
% \affiliation{Hefei National Laboratory for Physical Sciences at the Microscale, University of Science \& Technology of China, Hefei 230026, People’s Republic of China}
\affiliation{School of Emerging Technology, University of Science and Technology of China, Hefei 230026, People’s Republic of China}
\author{S.~M.~Gunasekera} % surani04@hotmail.com
\affiliation{Department of Physics, University of Bath, Claverton Down, Bath BA2 7AY, United Kingdom}
\author{C.~Wen} % surani04@hotmail.com
\affiliation{School of Physical Science and Technology, ShanghaiTech University, Shanghai 201210, People’s Republic of China}
\author{Y.~Zhang} % surani04@hotmail.com
\affiliation{School of Physical Science and Technology, ShanghaiTech University, Shanghai 201210, People’s Republic of China}
\author{D.~Prabhakaran} % dharmalingam.prabhakaran@physics.ox.ac.uk
\affiliation{Department of Physics, University of Oxford, Oxford, OX1 3PU, United Kingdom}
\author{F.~Flicker} % flicker@cardiff.ac.uk
\affiliation{School of Physics and Astronomy, Cardiff University, Cardiff CF24 3AA, United Kingdom}
\author{Y.~Chen} % yulin.chen@physics.oxford.ac.uk; 
\affiliation{Department of Physics, University of Oxford, Oxford, OX1 3PU, United Kingdom}
\affiliation{School of Physical Science and Technology, ShanghaiTech University, Shanghai 201210, People’s Republic of China}
\affiliation{CAS-Shanghai Science Research Center, Shanghai 201210, People’s Republic of China}
\author{M.~Mucha-Kruczy\'{n}ski}
\email{M.Mucha-Kruczynski@bath.ac.uk}
\affiliation{Department of Physics, University of Bath, Claverton Down, Bath BA2 7AY, United Kingdom}
\affiliation{Centre for Nanoscience and Nanotechnology, University of Bath, Claverton Down, Bath BA2 7AY, United Kingdom}

\begin{abstract}
We report on the interplay between a van Hove singularity and a charge density wave state in 2H-TaSe$_{2}$. We use angle-resolved photoemission spectroscopy to investigate changes in the Fermi surface of this material under surface doping with potassium. At high doping, we observe modifications which imply the disappearance of the $\threebythree$ charge density wave and formation of a different correlated state. Using a tight-binding-based approach as well as an effective model, we explain our observations as a consequence of coupling between the single-particle Lifshitz transition during which the Fermi level passes a van Hove singularity and the charge density order. \added{In this scenario,} the high electronic density of states associated with the van Hove singularity induces a change in the periodicity of the charge density wave from the known $\threebythree$ to a new $\twobytwo$ superlattice. 
\end{abstract}

\maketitle

\section{Introduction}

Because of the fundamental importance of the electrons in the vicinity of the Fermi surface (FS) for low-energy excitations, the shape of this surface has a significant impact on the properties of metals \cite{kaganov_spu_1979}. This is particularly evident when, as a function of some external parameter like pressure \cite{chu_prb_1970, godwal_prb_1998, nishimura_prl_2019}, temperature \cite{wu_prl_2015, zhang_natcomms_2017, chen_prl_2020}, magnetic field \cite{kozlova_prl_2005, rourke_prl_2008, orlita_prl_2012} or doping \cite{okamoto_prb_2010, norman_prb_2010, leboeuf_prb_2011}, the FS undergoes a change of topology resulting in a Lifshitz transition (also known as electronic topological transition) \cite{lifshitz_jetp_1960}. In contrast to the more conventional phase transitions described by the Landau theory, Lifshitz transitions do not involve symmetry breaking but still lead to singularities in many observables \cite{blanter_physrep_1994} because changes of topology of the equi-energetic surface are accompanied by van Hove singularities in the electronic density of states (DoS).

The impact of van Hove singularities is especially significant in low dimensions, $d\leq 2$, where divergences in the DoS are possible at some of the dispersion critical points. This is the case for saddle points for $d=2$ which lead to a logarithmic divergence in the DoS \cite{vanhove_physrev_1953, blanter_physletta_1994}, the presence of which is often implicated in promoting new orders in two-dimensional and layered materials \cite{rice_prl_1975, markiewicz_jpcs_1997, yudin_prl_2014, kim_prb_2018, hlubina_prl_1997, benhabib_prl_2015, braganca_prl_2018, wu_prl_2021, jiang_natmater_2021, hu_natcomms_2022, kang_natphys_2022, cao_nature_2018, nandkishore_natphys_2012, rosenzweig_prl_2020, varlet_prl_2014, varlet_synthmet_2014, zhou_nature_2021}, including high-temperature superconducting cuprates \cite{markiewicz_jpcs_1997, benhabib_prl_2015, braganca_prl_2018}, topological Kagome superconductors \cite{wu_prl_2021, jiang_natmater_2021, hu_natcomms_2022, kang_natphys_2022}, and magic-angle twisted bilayer graphene and other graphene materials \cite{cao_nature_2018, nandkishore_natphys_2012, rosenzweig_prl_2020, varlet_prl_2014, varlet_synthmet_2014, zhou_nature_2021}.

Here, we study the impact of a Lifshitz transition and the associated van Hove singularity on a charge density wave (CDW) -- a correlated ordering of electrons which form a standing wave pattern accompanied by a periodic distortion of the atomic lattice -- by surface doping bulk 2H-TaSe$_{2}$ with potassium. Using angle-resolved photoemission spectroscopy (ARPES), we map out directly the electronic dispersion in its low-temperature commensurate $\threebythree$ CDW, that is, one with a superstructure described by tripling of the in-plane primitive lattice vectors of the uncorrelated state. We then observe how it changes as the previously unoccupied electronic states in the topmost layers are filled so that the chemical potential crosses a saddle point in the dispersion. Based on calculations of generalized susceptibility within a minimal two-band model and effective description of the coupling between the saddle points, we conclude that the change in FS topology drives a change in the CDW from a $\threebythree$ to a $\twobytwo$ order. \added{Such a scenario supports} a theoretical prediction from almost half a century ago \cite{rice_prl_1975} which suggested involvement of the van Hove singularities in the formation of the CDW in TaSe$_{2}$ but was later shown not to be relevant for the $\threebythree$ phase. It also demonstrates the potential of engineering many-body phases using van Hove singularities. 

\section{Topology of electronic bands of 2H tantalum diselenide}

%%%%% Fig - lattice and bands %%%%%
\begin{figure}[t]
  \centering
  \includegraphics[width=1.0\columnwidth]{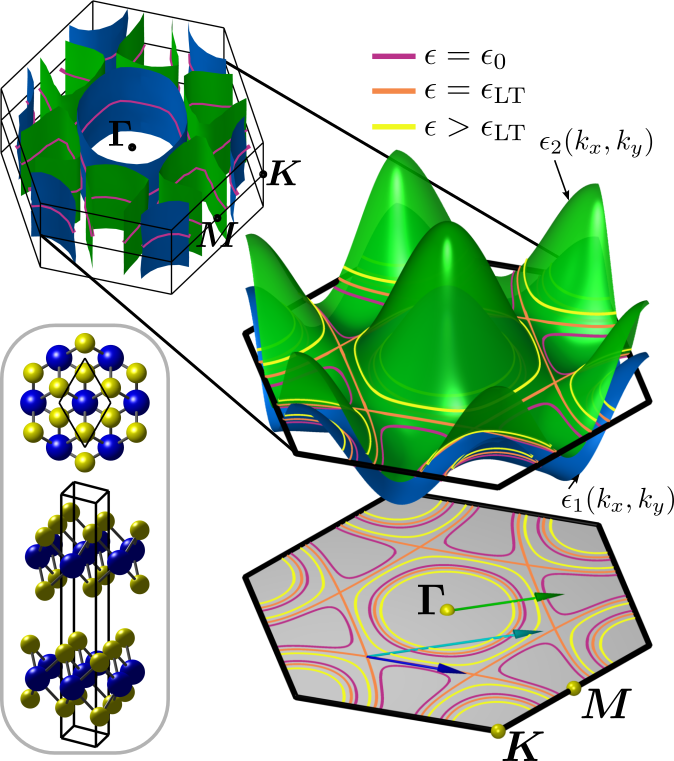}
  \caption{Dispersion and topology of electronic bands of 2H-TaSe$_{2}$. In the top left, the quasi-2D Fermi surface of undoped 2H-TaSe$_{2}$ is shown within the bulk Brillouin zone (hexagonal prism), as calculated with density functional theory (see Appendix~\ref{sec:ab_initio}), with the blue and green surfaces marking two different bands. The change of the dispersion and topology of these two bands with doping is shown in the centre for $k_{z}=0$ for $\epsilon=\epsilon_{0}$ (purple contours; also shown within the bulk Brillouin zone) as well as for $\epsilon=\epsilonLT=\epsilon_{0}+0.05\,\mathrm{eV}$ (orange) and $\epsilon>\epsilonLT$ (yellow). The same contours are shown again on top of the grey regular hexagon, representing the projected two-dimensional Brillouin zone, below the dispersion. Also shown are arrows depicting representative $\qvector$ wave vectors for the $\threebythree$ [green], $\twobytwo$ [cyan] and $\sqrtthree$ [blue] charge density wave orders. The inset in bottom left shows the top and side views of 2H-TaSe$_{2}$ lattice structure. Ta and Se atoms are marked with blue and yellow spheres, respectively, and the right rhombic prism indicated with the black solid lines is the unit cell.}
  \label{fig:bands}
\end{figure}
%%%%% End of Fig - lattice and bands %%%%%

2H-TaSe$_{2}$ consists of weakly coupled layers each of which is made of a plane of tantalum atoms sandwiched between two planes of seleniums. The consecutive layers are rotated by 180$^{\circ}$ and stacked so that the transition metals are placed on top of each other, as shown in the inset in the bottom left of Fig.~\ref{fig:bands}. The material exhibits a second-order transition into a near commensurate $\threebythree$ CDW phase at 122 K, followed by a first-order transition which locks the charge order into the $\threebythree$ superlattice at 90 K \cite{wilson_advphys_1975, rossnagel_jpcm_2011}. As suggested by the high critical temperature of the transition into the commensurate CDW, electronic band reconstruction in this phase is quite strong, with the CDW gap $~50$--100 meV in high quality crystals \cite{rossnagel_prb_2005, li_prb_2018}. For this reason, 2H-TaSe$_{2}$ serves as a model to understand the electronic properties of the isostructural and isoelectronic 2H-NbSe$_{2}$, 2H-TaS$_{2}$ and 2H-NbS$_{2}$ in which charge order is weaker so that only the incommensurate phase appears \cite{naito_jpsj_1982, moncton_prb_1977} (in NbS$_{2}$ the CDW order is so fragile that it has only been observed in its two-dimensional limit \cite{lin_nanores_2018, bianco_nanolett_2019}) as well as the mechanisms behind CDW phases in general. 

% Describe the evolution of thinking about CDW:

The driving force behind the $\threebythree$ CDW was long debated, partly due to incorrect predictions of the position of the Fermi level \cite{wilson_advphys_1975, wexler_jpc_1976, wilson_prb_1977, castroneto_prl_2001}. Experimental and computational studies have established that the Fermi surface, shown in the top left of Fig.~\ref{fig:bands}, consists of $\vect{\Gamma}$- and $\vect{K}$-centered tube-shaped hole sheets from the first band (blue) and $\vect{M}$-centered electron “dogbone” sheets from the second (green) and that the CDW is driven by a combination of Fermi surface nesting and electron-phonon coupling \cite{rossnagel_prb_2005, inosov_njp_2008, laverock_prb_2013, li_prb_2018}. Because of the quasi-two-dimensional nature of the FS, in what follows we focus on the plane for which the out-of-plane component of the wave vector, $k_{z}$, is constant (we choose $k_{z}=0$). This allows us to parametrize the two relevant bands as surfaces with energy, $\epsilon_{i}\equiv\epsilon_{i}(k_{x},k_{y})$, ($i=1,2$), dependent on in-plane wave vector, $\vect{k}=(k_{x},k_{y})$, and shown in the right of Fig.~\ref{fig:bands} using the same colours, green and blue, as for the bulk FS. The FS then becomes a Fermi contour (FC), indicated in purple solid lines on top of the full FS in top left as well as on top of the dispersion surfaces $\epsilon_{i}$ and the gray regular hexagon below which represents the two-dimensional Brillouin zone of TaSe$_{2}$. We denote the Fermi energy of the bulk as $\epsilon=\epsilon_{0}$. 

By inspecting the green dispersion surface, $\epsilon_{2}$, it can be seen that an increase in the Fermi energy leads to a change of the topology of the Fermi contour. As the energy increases above $\epsilon_{0}$, the $\vect{\Gamma}$-pocket decreases slightly. At the same time, the dogbone pockets grow and connect with each other at the energy $\epsilon_{\mathrm{LT}}\approx\epsilon_{0}+0.05\,\mathrm{eV}$ and momentum close to $\tfrac{1}{2}\vect{K}$, which determines the position of saddle points of the green surface, leading to energy contours as shown in orange in Fig.~\ref{fig:bands}. For energies $\epsilon>\epsilon_{\mathrm{LT}}$, the connected dogbones split to form another set of $\vect{K}$-centred pockets as well as one more centred around $\vect{\Gamma}$ (contours shown in yellow). Our work is motivated by the presence of this saddle point, with the potential to tune the Fermi level through a Lifshitz transition -- change in the topology of the Fermi contour from the three dogbone pockets to one $\vect{\Gamma}$- and one $\vect{K}$-centred pockets -- and the question of how it impacts the CDW order.  

\section{ARPES spectra of surface doped 2H-{T\lowercase{a}S\lowercase{e}}$_{2}$}

%%%%% Fig - ARPES maps %%%%%
\begin{figure*}[t]
  \centering
  \includegraphics[width=0.97\textwidth]{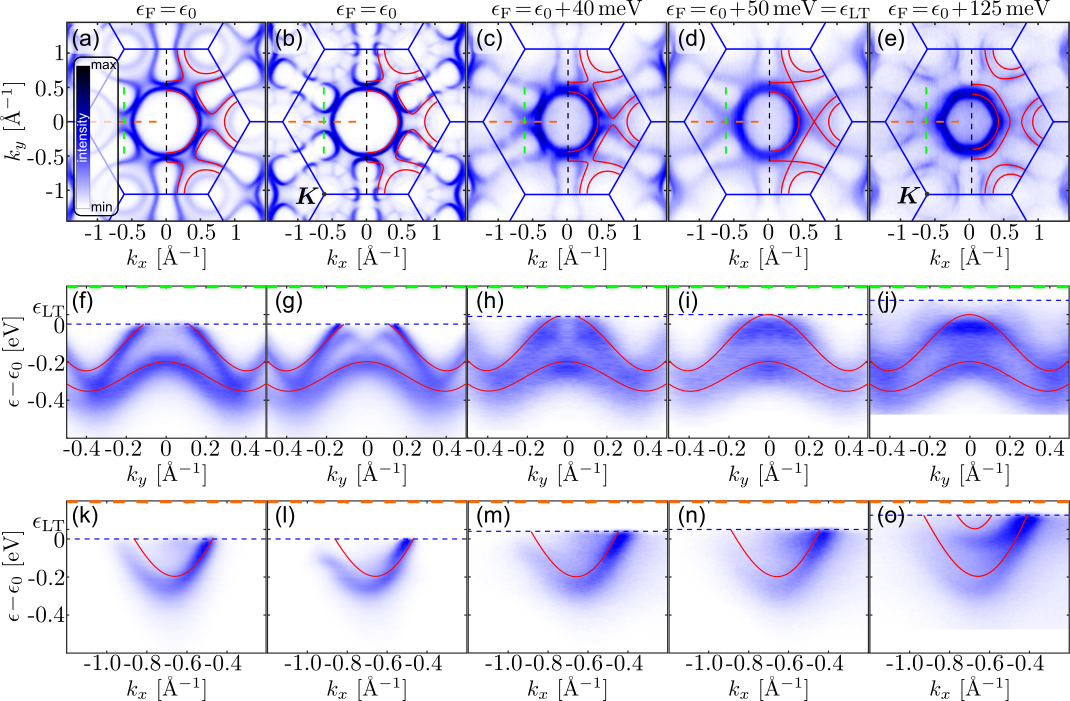}
  \caption{\added{Angle-resolved photoemission spectra of 2H-TaSe$_{2}$: (a), (f), (k) as grown and at the temperature $T=130\,$K; (b), (g), (l) at the temperature $T=10\,$K and without surface doping; (c), (h), (m) after the first, (d), (i), (n) second and (e), (j), (o) third dose of surface potassium deposition at $T=22\,$K. The top row [panels (a)-(e)] shows constant-energy maps at the Fermi level; the middle row [panels (f)-(j)] shows spectra along the path crossing the saddle point at $\tfrac{1}{2}\vect{K}$ as indicated in green dashed lines in panels (a)-(e); the bottom row [panels (k)-(o)] shows spectra along the path crossing the saddle point as indicated in orange dashed lines in panels (a)-(e). The blue solid lines indicate the Brillouin zone boundaries and the red solid lines show bands as predicted by our tight-binding model. The position of the Fermi energy, $\epsilon_{\mathrm{F}}$, for each case is provided above the constant-energy maps and indicated with the dashed blue line in panels (f)-(o). ARPES spectra were symmetrized with respect to $k_{x}=0$ and $k_{y}=0$ for better comparisons with calculations (see the SM for a brief discussion of this process \cite{supplement}); they were all plotted using the same intensity scale with colour mapping as shown in the inset of panel (a).}}
  \label{fig:arpes_maps}
\end{figure*}
%%%%% End of Fig - ARPES maps %%%%%

In order to study the electronic band structure of 2H-TaSe$_{2}$, we use angle-resolved photoemission spectroscopy (ARPES). Our measurements have been taken at the Diamond Light Source (I05) and Advanced Light Source (MERLIN) using photons with energy of $80\,\mathrm{eV}$ (see the Supplemental Material (SM) \cite{supplement} for further details of the ARPES measurements). The map of the Fermi surface as measured at the temperature $T=130\,$K is shown in Fig.~\ref{fig:arpes_maps}(a), clearly reflecting the single-particle Fermi contour at $\epsilon=\epsilon_{0}$ shown in purple in Fig.~\ref{fig:bands} (for direct comparison, in all panels of Fig.~\ref{fig:arpes_maps} we draw in solid red lines the theoretical band cuts obtained using the \added{single-particle tight-binding} model discussed in Sec.~\ref{sec:theory}). When measured at the temperature $T=10\,$K, Fig.~\ref{fig:arpes_maps}(b), the Fermi surface undergoes reconstruction due to the formation of the $\threebythree$ charge density wave \cite{rossnagel_prb_2005, borisenko_prl_2008, laverock_prb_2013, li_prb_2018}: (i) the circular pockets around $\vect{K}$ are gapped; instead, a set of six triangular features appears around each Brillouin zone corner; (ii) the dogbones become more rounded and develop gaps along their contour due to Bragg scattering caused by the CDW periodicity; (iii) the $\vect{\Gamma}$ pocket remains unaffected. \added{Because the CDW super-potential is much weaker than the original periodic lattice potential, the spectral intensity of CDW-folded bands is much weaker than the original, ``main'' bands. This allows us to track the latter and determine that the connectivity between the dogbones remains unchanged upon CDW formation so that the presence of a Lifshitz transition can still be anticipated if the Fermi energy could be increased.}   

In order to tune the Fermi energy, we deposit potassium atoms on the surface of our samples at the temperature of $22\,$K. The small electron affinity of potassium makes it a strong electron donor to most surfaces and leads to $n$-doping of the top layers of the crystal (the low temperature makes intercalation unfeasible). At the same time, surface sensitivity of ARPES means that only the very top layers are probed experimentally \cite{riley_nphys_2014, beaulieu_prl_2020, king_chemrev_2021} (2H-TaSe$_{2}$ unit cell height is $c=1.228$ nm \cite{yan_scirep_2015} as compared to the electron escape depth $l\sim 1$ nm \cite{seah_sia_1979}). In Fig.~\ref{fig:arpes_maps}(c)-(e), we show the Fermi surfaces as measured after three consecutive potassium depositions (the position of the Fermi energy, $\epsilon_{\mathrm{F}}$, after each deposition is estimated based on the effective model described in Sec.~\ref{sec:theory}). After the first potassium dose, panel (c), spectral features around the $\vect{K}$ points disappear while the most intense signal is from the states around $\vect{\Gamma}$. Importantly, connectivity between dogbones remains the same as in panels (a) and (b), implying that the Fermi level is still below the dispersion saddle points. Moreover, similar gaps along a dogbone contour can be identified in panels (b) and (c), indicating persistence of the $(3\times 3)$ CDW. Following a second potassium dose, panel (d), broadening of the dogbones indicates that the Fermi level is in the close vicinity of the saddle points. While it is difficult to determine whether the change of topology has already occured at that point, after the third potassium dose and an additional shift of the Fermi level, panel (e), connectivity of the Fermi surface has changed: merging of the dogbones leads to the formation of a circular pocket around $\vect{K}$ as well as one around $\vect{\Gamma}$. \added{Note that weak intensity features connect the pocket around $\vect{\Gamma}$ to pockets around $\vect{K}$ -- we assign these to the new order we discuss in the rest of the text.}

To further confirm that after the third dose of potassium the Fermi level moved above the saddle point, for each of the panels (a) to (e), we show in (f) to (j) the measured band dispersion along the momentum path perpendicular to the $\vect{\Gamma}$-$\vect{K}$ direction and passing through the location of the saddle point as indicated with the green dashed lines. \added{In panels (f)-(h), the higher energy band is not entirely below the Fermi level [note that the weaker intensity, M-shaped band in panel (g) is due to the $\threebythree$ CDW \cite{li_prb_2018}].} In contrast, in panels (i) and (j), this band is fully below the Fermi level which demonstrates that the Lifshitz transition has occurred. \added{The same can be seen in the bottom row of, panels (k)-(o), in which we show the measured band dispersion along the $\vect{K}-\vect{\Gamma}$ direction passing through the saddle point. This is the direction perpendicular to that shown in panels (f)-(j) and along which the dispersion displays opposite curvature. In panels (n) and (o), showing dispersion after the second and third potassium deposition, respectively, one can identify spectral features indicating another band in the vicinity of the Fermi level. This again implies that the Fermi level has crossed the saddle point (we comment further on the ARPES maps and the Lifshitz transition in the SM).}

\begin{figure}[tb]
\centering
\includegraphics[width=1.00\columnwidth]{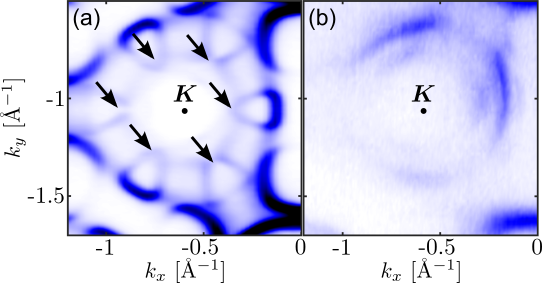}
\caption{Close-up on the ARPES Fermi level maps for 2H-TaSe$_{2}$ in the vicinity of the Brillouin zone corner $\vect{K}$ [as marked in Fig.~\ref{fig:arpes_maps}(b) and (e)], (a) before surface doping with potassium and (b) after the third dose of potassium deposition. Black arrows mark the signature of the C6 symmetry in panel (a). The spectra have been taken at the temperature $T=10\,$K and $T=22\,$K for panel (a) and (b), respectively, below the transition temperature of the commensurate $\threebythree$ charge density wave.}
\label{fig:symmetry}
\end{figure}

The spectra in \added{all but the first column of Fig.~\ref{fig:arpes_maps}} have been taken at a temperature significantly below the transition temperature of the commensurate $\threebythree$ CDW, $T_{c}\approx 90\,$K \cite{wilson_advphys_1975, rossnagel_jpcm_2011}. This allows us to study the interplay between the Lifshitz transition, driven by the single-particle electronic band structure, and the charge density order present in the material. In Fig.~\ref{fig:symmetry}, we compare the ARPES Fermi level maps in the vicinity of the Brillouin zone corner $\vect{K}$ for pristine 2H-TaSe$_{2}$ surface \added{in the $\threebythree$ state}, (a), as well as the surface after the final deposition of potassium, (b) [zoom in of parts of the maps in Fig.~\ref{fig:arpes_maps}(b) and (e), respectively]. \added{Within the single-particle picture, the symmetry of the band dispersion around $\vect{K}$ is $C_{3}$. In panel (a) of Fig.~\ref{fig:symmetry}, this is reflected in the three-fold symmetry of the outer band corresponding to the dark blue intensity contour (interrupted in some places by CDW-induced gaps). The inner band, however, displays a set of six weak-intensity features indicated with black arrows. These features are also due to the $\threebythree$ state, the wave vector of which folds the two otherwise inequivalent Brillouin zone corners onto each other \cite{li_prb_2018} and hence allows for a weak six-fold, rather than exclusively three-fold, symmetry \cite{rossnagel_prb_2005, li_prb_2018}. After the third dose of potassium, panel (b), the $C_{6}$ features in the inner band disappear and only the $C_{3}$ symmetry remains. Because the Lifshitz transition alone cannot be responsible for this change in symmetry, we conclude that the electronic order has changed as a result of the Fermi level shift and the $\threebythree$ CDW is no longer present. At the same time, modulation of the photoemission intensity along the outer and inner bands in Fig.~\ref{fig:symmetry}(b) is indicative of new gaps in the electronic dispersion incompatible with the single-particle picture. We compare the intensity profiles along the inner band before and after potassium doping in the SM.}

%%%%% Fig - ARPES maps %%%%%
%%%%% End of Fig - ARPES maps %%%%%

\section{Discussion}\label{sec:theory}

In order to understand the impact of doping and the Lifshitz transition on the charge density order, we describe our system using an effective two-band model based on the tight-binding expansion due to the Ta sites only (as the transition metal $d$-orbitals provide a predominant contribution to the bands crossing the Fermi level) \cite{smith_jpc_1985, rossnagel_prb_2005, inosov_prb_2009, li_prb_2018}, with one orbital per site. Because of the quasi-two-dimensionality of the bulk band structure, we focus on the dispersion for a constant $k_{z}$. For simplicity, we choose $k_{z}=0$ -- the exact choice is, however, not important as the functional form of the model is independent of $k_{z}$ and its parameters are prescribed by experimental data (we comment further on the importance of the out-of-plane dispersion in the SM \cite{supplement}). The essential features of the band structure can be reproduced by keeping terms up to next-nearest intra- and interlayer neighbours,
\begin{align}\label{eqn:hamiltonian} %\op{H}_{\mathrm{TB}}
  & \HTB(\vect{k}) = f(t_0,t_{1},t_{2};\vect{k})\sigma_{0}+f(\tilde{t}_{0},\tilde{t}_{1},\tilde{t}_{2};\vect{k})\sigma_{x},\\
  % f_{1}(\vect{k}) = & \epsilon, \\
  & f(\alpha,\beta,\gamma;\vect{k}) = \alpha+2\beta\left[ \cos k_{x}a + 2\cos\frac{k_{x}a}{2}\cos\frac{\sqrt{3}k_{y}a}{2} \right] \nonumber \\
  & +2\gamma\left[ \cos\sqrt{3}k_{y}a + 2\cos\frac{3k_{x}a}{2}\cos\frac{\sqrt{3}k_{y}a}{2} \right], \nonumber
\end{align}
where $\sigma_{0}$ is a $2\times 2$ identity matrix, $\sigma_{x}$ is the $x$ Pauli matrix, $a=3.43$ {\AA} is the lattice constant of 2H-TaSe$_{2}$, $t_0=0.113\,\mathrm{eV}$ is the Ta on-site energy, $\tilde{t}_{0}=0.184\,\mathrm{eV}$ is the direct interlayer coupling and $t_{1}=0.073\,\mathrm{eV}$ ($\tilde{t}_{1}=0.029\,\mathrm{eV}$) and $t_{2}=0.142\,\mathrm{eV}$ ($\tilde{t}_{2}=0.038\,\mathrm{eV}$) are the nearest and next-nearest intralayer (interlayer) couplings. We fixed our parameters using a hybrid approach in which we fit the model to the ARPES data below the Fermi energy and density functional theory calculations above (see Appendix~\ref{sec:ab_initio} for the details of the latter). We have also tuned the on-site term $t_0$ so that $\epsilon_{\mathrm{LT}}=0$. Our model provides the minimal description which captures the topology of the Fermi contour as a function of the Fermi level and allows us to investigate analytically the saddle points located at $\vect{k}^{(n)}_{\mathrm{LT}}=\op{R}_{n\pi/3}[\tfrac{2\pi}{3a}+\tfrac{4}{\sqrt{5}a}\delta,0]^{T}$, $n=0,1,\ldots,5$, where $\op{R}_{\theta}$ is the operator of rotation by angle $\theta$ and $\delta=\tfrac{t_{1}+\tilde{t}_{1}}{6(t_{2}+\tilde{t}_{2})}\ll 1$. While we find that introducing further neighbours or including nonorthogonality corrections allows to fit the band structure better, the additional parameters complicate the description of the saddle point without providing new insight into the physics.

Following the approach developed in Ref.~\cite{varma_prl_1977, varma_prb_1979} and previously applied for example to 2H-NbSe$_{2}$ \cite{flicker_natcomm_2015, flicker_prb_2016}, we use the Hamiltonian $\HTB(\vect{k})$ to approximate the electron-phonon matrix element, $g^{i}_{\vect{k},\vect{k}+\vect{q}}$, which describes scattering of an electron in band $i$ from a state with wave vector $\vect{k}$ to a state in the same band with wave vector $\vect{k}+\vect{q}$ with the simultaneous absorption (emission) of a phonon with wave vector $\vect{q}$ ($-\vect{q}$), using gradients of the electronic dispersion,
\begin{align}
  g^{i}_{\vect{k},\vect{k}+\vect{q}}=\left[\nabla\epsilon_{i}(\vect{k}+\vect{q})-\nabla\epsilon_{i}(\vect{k})\right]\cdot\frac{\vect{q}}{|\vect{q}|},
\end{align}
where we have ignored a constant prefactor. Moreover, because formation of the charge density wave in 2H-TaSe$_{2}$ involves softening of the longitudinal acoustic phonon \cite{moncton_prb_1977}, we have projected the electron-phonon coupling on the direction of momentum transfer. Also, in the scheme of Ref.~\cite{varma_prl_1977, varma_prb_1979}, the symmetric form of the Hamiltonian $\HTB$ in Eq.~\eqref{eqn:hamiltonian} implies that the interband electron-phonon coupling is strictly zero. 

%%%%% Fig - ARPES maps %%%%%
\begin{figure}[tb]
\centering
\includegraphics[width=1.00\columnwidth]{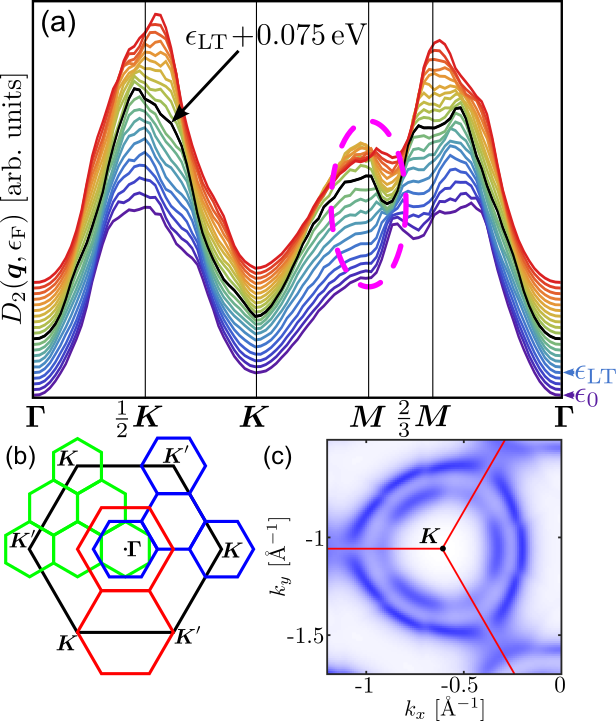}
\caption{(a) Generalized static susceptibility $D_{2}(\vect{q},\epsilon_{\mathrm{F}})$ as a function of the Fermi level, $\epsilon_{\mathrm{F}}$, from $\epsilon_{\mathrm{F}}=\epsilon_{0}$ (purple) to $\epsilon_{\mathrm{F}}=\epsilon_{0}+0.25\,\mathrm{eV}$ (red) in steps of $0.0125\,$eV, along the high-symmetry Brillouin zone directions. The curves have been shifted vertically for clarity. On the right, we indicate the curves corresponding to the Fermi energy of the pristine material, $\epsilon_{\mathrm{F}}=\epsilon_{0}$, as well as for the Fermi energy at the saddle point, $\epsilon_{\mathrm{F}}=\epsilon_{\mathrm{LT}}$. The curve for $\epsilon_{\mathrm{F}}=\epsilon_{0}+0.125\,\mathrm{eV}=\epsilon_{\mathrm{LT}}+0.075\,\mathrm{eV}$, corresponding to the Fermi level estimate for Fig.~\ref{fig:arpes_maps}(e) and (j), is shown in black. The bright purple dashed oval highlights a susceptibility feature at $\vect{M}$ which appears in the energy range $\sim 0.1\,\mathrm{eV}$ above $\epsilon_{\mathrm{LT}}$. (b) Comparison of Brillouin zones of the uncorrelated phase (black) and the $\threebythree$ (blue), $\sqrtthree$ (green) and $\twobytwo$ (red) superlattices. (c) Spectral function in the vicinity of $\vect{K}$ computed at the energy $\epsilon_{0}+0.125\,\mathrm{eV}$ for the $\twobytwo$ CDW using the model described in Appendix \ref{sec:appendix} with broadening $\eta=0.1\,\mathrm{eV}$. The red lines depict the boundaries of the $\twobytwo$ superlattice Brillouin zone.}
\label{fig:susceptibility}
\end{figure}
%%%%% End of Fig - ARPES maps %%%%%

The knowledge of the electron-phonon coupling allows us to compute the static generalized susceptibility,
\begin{align}
  D_{2}(\vect{q},\epsilon_{\mathrm{F}})=\sum_{i}\int_{BZ}d\vect{k}\left( g^{i}_{\vect{k},\vect{k}+\vect{q}} \right)^{2}\frac{f\left[\epsilon_{i}(\vect{k})\right]-f\left[\epsilon_{i}(\vect{k}+\vect{q})\right]}{\epsilon_{i}(\vect{k})-\epsilon_{i}(\vect{k}+\vect{q})},
\end{align}
where, for a given Fermi energy, $\epsilon_{\mathrm{F}}$, $f[\epsilon_{i}(\vect{k})]$ is the filling factor of the state with energy $\epsilon_{i}(\vect{k})$ in band $i$ and with wave vector $\vect{k}$, and the integral is over the two-dimensional Brillouin zone. As peaks of $D_{2}(\vect{q},\epsilon_{\mathrm{F}})$ provide information about structural instabilities in the material \cite{varma_prl_1977, varma_prb_1979, heil_prb_2014}, we present the plots of $D_{2}(\vect{q},\epsilon_{\mathrm{F}})$ for the high-symmetry directions $\vect{\Gamma}$--$\vect{K}$--$\vect{M}$--$\vect{\Gamma}$ in Fig.~\ref{fig:susceptibility}(a). The curves for increasing $\epsilon_{\mathrm{F}}$ have been shifted for clarity, starting from $\epsilon_{\mathrm{F}}=\epsilon_{0}$ (purple) to $\epsilon_{\mathrm{F}}=\epsilon_{0}+0.25\,\mathrm{eV}$ (red). Two features persist across the whole range of energies: (i) a broad peak in the vicinity of $\tfrac{2}{3}\vect{M}$ and (ii) a peak in the vicinity of $\tfrac{1}{2}\vect{K}$. The former is related to the $\threebythree$ CDW \cite{inosov_njp_2008, laverock_prb_2013}. Given the symmetry of the ARPES constant-energy maps which excludes considerations of one-dimensional CDW, the latter would imply a $\sqrtthree$ order with a wavevector that nests the saddle points onto each other, as shown with the blue arrow in the bottom right of Fig.~\ref{fig:bands} (ideal nesting occurs for $\delta\rightarrow 0$). Interestingly, another feature, strongly dependent on $\epsilon_{\mathrm{F}}$, appears in $D_{2}(\vect{q},\epsilon_{\mathrm{F}})$ at $\vect{M}$ as highlighted in Fig.~\ref{fig:susceptibility} by the bright purple dashed oval. This lower peak which develops for $\epsilon_{\mathrm{F}}$ slightly above $\epsilon_{\mathrm{LT}}$ [we show in black the curve for $\epsilon_{\mathrm{F}}=\epsilon_{\mathrm{LT}}+0.075\,\mathrm{eV}$ which corresponds to the Fermi level estimate for Fig.~\ref{fig:arpes_maps}(e) and (j)] but disappears for $\epsilon_{\mathrm{F}}\gtrsim\epsilon_{\mathrm{LT}}+0.15\,\mathrm{eV}$ suggests potential instability towards a $\twobytwo$ CDW. Such order would also nest saddle points onto each other as shown with the cyan arrow in Fig.~\ref{fig:bands} (again, ideal nesting takes place for $\delta\rightarrow 0$), albeit in two groups of three. 

%Given the effective nature of our model, we do not rely on the relative heights of the $D_{2}(\vect{q},\epsilon_{\mathrm{F}})$ features to identify the leading instability. Instead, 

%%%%% Fig - ARPES maps %%%%%
\begin{figure*}[t!]
  \centering
  \includegraphics[width=1.00\textwidth]{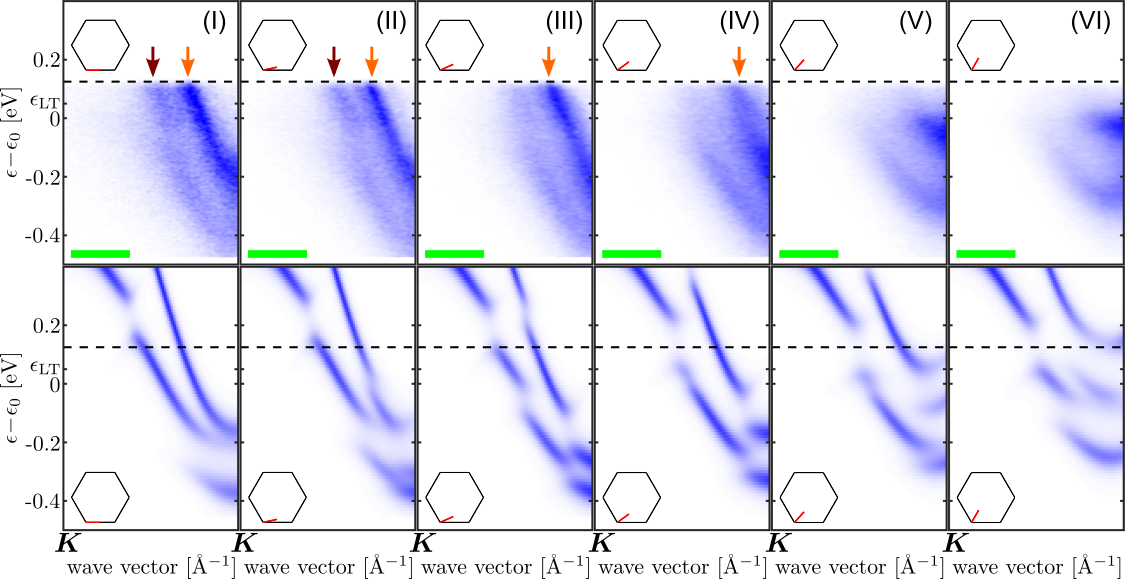}
  \caption{Comparison of experimental (top) and simulated (bottom) ARPES intensities following the final potassium deposition for wave vector cuts at different angles starting from the $\vect{K}$ point of the Brillouin zone (as indicated with the red lines in the insets of each panel). The Fermi energy, $\epsilon_{\mathrm{F}}=\epsilon_{0}+0.125\,\mathrm{eV}$, is indicated with the dashed black line. In the top row, the dark red and orange arrows indicate the momentum at which the inner and outer band, respectively, cross the Fermi level. All cuts cover the same distance in $k$-space; the green scale bars in the top row correspond to 0.2 \AA$^{-1}$. For the theoretical plots, we used energy broadening $\eta=0.05\,\mathrm{eV}$.}
  \label{fig:comparison}
\end{figure*}
%%%%% End of Fig - ARPES maps %%%%%

Motivated by the symmetry breaking shown in Fig.~\ref{fig:symmetry}, we compare in Fig.~\ref{fig:susceptibility}(b) the Brillouin zones of the uncorrelated phase (black) as well as the $\threebythree$ (blue), $\sqrtthree$ (green) and $\twobytwo$ (red) CDW. Note that for both the $\threebythree$ and $\sqrtthree$ superlattices, the two originally inequivalent zone corners $\vect{K}$ and $\vect{K}'$ are folded onto $\vect{\Gamma}$. In contrast, they remain inequivalent for the $\twobytwo$ superlattice, guaranteeing $C_{3}$ symmetry of dispersions in their vicinity in agreement with Fig.~\ref{fig:symmetry}(b). In Fig.~\ref{fig:susceptibility}(c), we show the spectral function in the vicinity of the corner $\vect{K}$, computed at the energy $\epsilon_{0}+0.125\,\mathrm{eV}$ for the $\twobytwo$ CDW using the approach described in Appendix \ref{sec:appendix}. In agreement with Fig.~\ref{fig:symmetry}(b), each band is divided into three arcs, with the spectral weight decreasing where the contours cross the red lines which indicate the boundaries of the $\twobytwo$ Brillouin zone. This suggests that gaps observed in Fig.~\ref{fig:symmetry}(b) can be understood as a consequence of Bragg scattering of electrons by the newly formed $\twobytwo$ superlattice. We show in the SM \cite{supplement} that similar spectral function maps for the $\threebythree$ and $\sqrtthree$ phases display features which disagree with our observations.

While both the $\sqrtthree$ and $\twobytwo$ phases lead to nesting of the saddle points with each other, further support for the latter comes from studying a minimal model describing such coupling which we present in Appendix \ref{sec:LT}. For the $\twobytwo$ phase, introducing attractive coupling between electrons in the vicinity of the saddle points leads to opening of a gap at a filling corresponding to the position of the Fermi level at the saddle point for the uncorrelated state. This agrees with experimental observations -- gap above the saddle point in Fig.~\ref{fig:arpes_maps}(j). At the same time, no such gap opens in the $\sqrtthree$ phase. We have also confirmed that formation of a $\twobytwo$ phase\added{, as described by our effective model,} lowers the total electron energy of the system as compared to the uncorrelated state (see further discussion in the SM \cite{supplement}). This energetic benefit is due to the new phase opening gaps in the electronic spectrum which remove the van Hove singularities located at the saddle points from the vicinity of the Fermi level.

In Fig.~\ref{fig:comparison}, we show in the top row the measured ARPES intensities for the sample after the third deposition of potassium, for $k$-space cuts in different directions from $\vect{K}$: from $\vect{K}$ towards $\vect{M}$ (first panel) to $\vect{K}$ towards $\vect{\Gamma}$ (last panel), in constant intervals. These cuts provide further information on the evolution of the band gaps observed as discontinuities of the Fermi contour in Fig.~\ref{fig:symmetry}(b). We mark with maroon and orange arrows the wave vectors at which the two bands, \added{which correspond to the inner and outer contours, respectively, around $\vect{K}$ in Fig.~\ref{fig:symmetry}(b)}, cross the Fermi level, $\epsilon_{\mathrm{F}}=\epsilon_{0}+0.125\,\mathrm{eV}$ (the latter is indicated in all the panels by the black dashed line). In panel (III), the intensity of the inner band gradually fades away and the band exhibits a gap-like feature at the Fermi level. This feature persists in the remaining panels, (IV)-(VI). In turn, the outer band becomes flatter as the angle increases, suggestive of the formation of a gap in panels (V) and (VI) (the latter is equivalent to the $\vect{K}\!-\!\vect{\Gamma}$ direction). 

In the theoretical plots in the bottom row of Fig.~\ref{fig:comparison}, we have deliberately shown the intensity for the states above the Fermi level, to illustrate the evolution of the gaps in the outer and inner bands for different $k$-space cuts. Crucially, the trend of a gap at the Fermi level appearing first in the inner band, followed by a gap in the outer band, is preserved. In the inner band, this is due to a band gap moving down in energy as the direction of the cut moves from $\vect{K}\!-\!\vect{M}$ to $\vect{K}\!-\!\vect{\Gamma}$. In the outer band, the gap appears due to the coupling of the saddle points with each other and is hence maximised along the $\vect{K}\!-\!\vect{\Gamma}$ direction which passes through a saddle point. With the help of an effective model of coupled saddle points, (see Appendix~\ref{sec:LT}), this maximum gap can be estimated as $\sim 3\gaptwo$, with $\gaptwo$ the intraband CDW gap parameter. \added{While the data presented here were obtained using photons with linear horizontal polarization, we have performed measurements using linear vertical polarization as well (we present a comparison of the two in the SM \cite{supplement}). The agreement between the two polarizations and the theoretically calculated spectral function suggests that effects related to the polarization of the incoming light or matrix element effect \cite{kurtz_jpcm_2007, damascelli_rmp_2003} do not play any role in our observations. Also, we do not observe any novel band-like features in our spectra expected if the modifications of the dispersion are due to ordering of potassium on the surface of TaSe$_{2}$ (see \cite{supplement} for an additional discussion of potassium deposition).}

As the bulk of our crystal necessarily continues to host the $\threebythree$ charge density wave, it might be interesting to explore the crossover between the bulk and the surface correlated states, possibly using soft-X-ray ARPES with a longer photoelectron attenuation depth than traditional photoemission \cite{strocov_srn_2014}. At the same time, because the change of charge density order relies on nesting of electronic van Hove singularities, the $\twobytwo$ phase should be further stabilised in thin enough (with a number of layers $\sim 2$) doped or gated crystals of TaSe$_{2}$, similarly to the case of TaS$_{2}$ \cite{hall_acsnano_2019}. \added{Thin flakes also have the advantage that the Fermi level could be shifted using electrostatic gating rather than by surface deposition of alkali atoms as done here. In that case, scanning transmission microscopy could be used to observe the existence of the new CDW directly in real space (in contrast to our situation in which, as we discuss in the SM, the deposited potassium prevents access to the TaSe$_{2}$ surface). Another approach would be alloying of Ta with another metal. For example, tungsten atoms have one more $d$-electron than tantalum so that in Ta$_{1-x}$W$_{x}$Se$_{2}$ the Fermi level crosses the vHS for a small W concentration, $x$ \cite{wan_npj2d_2023}. Moreover, in 2H-NbSe$_{2}$, the Fermi level lies above a saddle point equivalent to the one discussed here and the material displays the incommensurate $\threebythree$ order \cite{inosov_njp_2008} [this matches up with the persistence of the peak corresponding to the $\threebythree$ instability in generalized susceptibility as a function of the Fermi level in Fig.~\ref{fig:susceptibility}(a)]. This means that in Ta$_{1-x}$Nb$_{x}$Se$_{2}$, the Fermi level has to cross the saddle point for some Nb concentration, $x$ (as Nb is isoelectronic with Ta, this is because of a deformation of the band structure which should occur smoothly between the two pure compositions, rather than due to electronic doping per se). In both cases, engineering of the band structure might allow to study the presence of the $\twobytwo$ state without obstructing the surface.
}

To mention, the presence of saddle points roughly midway between $\vect{\Gamma}$ and $\vect{K}$ in the $d$-orbital-derived bands is a generic feature of the band structure of all 2H transition metal dichalcogenides (including the semiconducting members of the family) and so we suggest that several seemingly unrelated observations of a $\twobytwo$ superlattice in these materials are connected by the underlying mechanism of van Hove singularity nesting. In the metallic TaSe$_{2}$, TaS$_{2}$, NbSe$_{2}$ and NbS$_{2}$, as discussed here, the relevant bands contribute to the Fermi surface and so the saddle points are relatively close to the Fermi level. Some spectroscopy measurements suggest a weak $\twobytwo$ charge density wave can coexist with the $\threebythree$ order in NbSe$_{2}$ \cite{chen_ssc_1984}. In the same work, no $\twobytwo$ superlattice was observed in TaSe$_{2}$ and TaS$_{2}$. However, this CDW order was observed in chalcogen poor TaSe$_{2}$ \cite{wilson_advphys_1975} which is consistent with the driving mechanism as discussed here given that chalcogen vacancies effectively $n$-dope the material \cite{rao_2dmater_2019, sayers_prm_2020, chee_afm_2020}. A $\twobytwo$ superlattice was also observed in intercalated TaSe$_{2}$ (and so highly $n$-doped), although the origin was suggested to be due to intercalant ordering rather than a new charge density order \cite{konig_epl_2012}. In semiconductors like MoS$_{2}$ or WSe$_{2}$, the saddle points can be found in the lowest lying conduction band. Moving the Fermi level into their vicinity requires a considerable doping, achievable for example by intercalation as suggested by the observations on MoS$_{2}$ \cite{subhan_nanolett_2021}.

\section{Summary}

To summarize, by depositing potassium on bulk 2H-TaSe$_{2}$, we have induced band bending near the surface of the crystal which allowed us to tune the Fermi level past a saddle point of the quasi-two-dimensional dispersion. We have used ARPES to observe the interplay between the resulting single-particle Lifshitz transition and the $\threebythree$ CDW existing in the material. With the help of an effective 2-band model fitted using both the ARPES data and density functional theory, we found that the resulting change in the Fermi surface is consistent with a change in CDW geometry from $\threebythree$ to $\twobytwo$. For the latter phase, spectral reconstruction nests saddle point van Hove singularities close to the Fermi level.

The saddle point nesting mechanism of CDW formation was proposed in 1975 \cite{rice_prl_1975}, shortly after the first experimental observations of CDWs in dimensions higher than one. Whereas in one-dimensional metals the weak-coupling nesting (Peierls) mechanism generically leads to CDW formation, in two and three dimensions a single wavevector will typically not connect large regions of the Fermi surface, and it was originally unclear how CDWs could be energetically beneficial. While saddle point nesting addressed this issue by taking advantage of the high density of states at the van Hove singularities, it is now accepted that in most transisition metal dichalcogenides CDWs are not generated by weak-coupling nesting instabilities, instead relying on the detailed structure of the interactions \cite{flicker_natcomm_2015, lin_natcomms_2020}. Our measurements on potassium-doped 2H-TaSe$_{2}$ could be the first observation of changes in the electronic dispersion due to the formation of saddle point-induced CDW order in transition metal dichalcogenides, as originally proposed.

Given the differences in mechanisms driving the $\threebythree$ and $\twobytwo$ CDW phases, it would be interesting to investigate how different their interplay with superconductivity is. For example, recently, a pair density wave (PDW) was observed in 2H-NbSe$_{2}$ using scanning Josephson tunnelling microscopy \cite{liu_science_2021}. In a PDW, it is Cooper pairs rather than electrons which break the symmetry of the crystal structure. This was the first evidence of a PDW outside the cuprate high-temperature superconductors and, owing to the remarkable similarity of the bandstructures of transition metal dichalcogenides, it would seem highly likely that a PDW could also be observed in 2H-TaSe$_{2}$ (albeit requiring temperatures below the critical temperature, $0.13\,$K for the pristine material \cite{kumakura_cjp_1996} in contrast to $7\,$K for NbSe$_{2}$ \cite{edwards_jpcs_1971}). In such a case, we might expect a similar change in PDW geometry under doping as discussed here for the CDW.

Finally, a doping-controlled change in CDW geometry from $\threebythree$ to $\twobytwo$ would constitute a quantum phase transition, indicating a presence of an underlying quantum critical point. A doping-based quantum phase transition is suspected to lead to the superconducting dome in the hole-doped cuprates \cite{broun_natphys_2008, michon_nature_2019}, and the opportunity to study a similar scenario in the absence of high-temperature superconductivity could help disentangle the complicated knot of intertwined phenomena accompanying that state. In the case of potassium-doped 2H-TaSe$_{2}$, the quantum phase transition would occur in the vicinity of a Lifshitz transition. The resulting high density of states makes the system highly tunable, leading to a change in the electron, crystal, and phonon structures. Access to such a controllable quantum phase transition would make doped 2H-TaSe$_{2}$ a promising candidate for future studies as a source of new exotic phenomena.

%- emphasize the attraction of 2D flakes which would not require potassium on top but can be electrostatically gated so that direct STM confirmation of our conclusions is possible
%- it is know that that NbSe2 has a Fermi level above the saddle point and also displays the $\threebythree$ order; our generalized susceptibility plots are consistent with this fact; studying an alloy of the two, Ta$_{1-x}$Nb$_{x}$Se$_{2}$ with Fermi level at the vHS should be possible 

\begin{acknowledgments}
We thank E.~Da Como and D.~Wolverson for their comments on the manuscript. This work has been supported by the UK Engineering and Physical Sciences Research Council (EPSRC) through the Centre for Doctoral Training in Condensed Matter Physics (CDT-CMP), Grant No. EP/L015544/1. Y.~L.~acknowledges support from the National Natural Science Foundation of China (Grant No.~12104304). J.~J.~acknowledges support from the National Natural Science Foundation of China (Grant No.~12174362). W.~R.~B.~L.~and Y.~L.~contributed equally to this work. \added{We also acknowledge beamtime allocations at the MERLIN beamline of the Advanced Light Source, USA, and the I05 beamline of the Diamond Light Source, UK, and support from beamline scientists J.~Denlinger, T.~Kim, and M.~Hoesch.} %The data used in this study are available from the University of Bath data archive at \href{https://doi.org/10.15125/BATH-00xxxxx}{https://doi.org/10.15125/BATH-00xxxxx}.

\end{acknowledgments}

\appendix

\section{\textit{ab initio} calculation of bandstructure}\label{sec:ab_initio}

In order to obtain the bulk band structure of 2H-TaSe$_{2}$ shown in the top left of Fig.~\ref{fig:bands}, we used the Quantum ESPRESSO package \cite{giannozzi_jpcm_2017} with relativistic pseudopotentials constructed using the PSLibrary \cite{dal_corso_cms_2014} for the local density approximation and a $10\times 10\times 10$ Monkhorst-Pack grid \cite{monkhorst_prb_1976} for the bulk crystal. From this, we extracted the data for $k_{z}=0$ plane (we show in the Supplemental Material \cite{supplement} that the dependence of the dispersion on $k_{z}$ is negligible) which we used together with the experimental data to fit our tight-binding model. Note that density functional theory calculations anticipate an incorrect position of the Fermi level. This is a well known effect in these materials \cite{laverock_prb_2013} which was corrected by comparison to the experiment.

\section{Modelling ARPES intensity of the $\twobytwo$ CDW}\label{sec:appendix}

We can simulate theoretically the experimental ARPES intensities by implementing Bragg scattering on a $\twobytwo$ superlattice, 
\begin{align}
\HCDW=\!\begin{bmatrix}
\HD(\vect{k}) & \gapmatrix & \gapmatrix & \gapmatrix \\
\gapmatrix & \HD(\vect{k}\!+\!\vect{G}_{0}) & \gapmatrix & \gapmatrix \\
\gapmatrix & \gapmatrix & \HD(\vect{k}\!+\!\vect{G}_{1}) & \gapmatrix \\
\gapmatrix & \gapmatrix & \gapmatrix & \HD(\vect{k}\!+\!\vect{G}_{2})
\end{bmatrix},
\end{align}
where $\HD(\vect{k})=\tfrac{1}{2}(\sigma_{x}+\sigma_{z})\HTB(\sigma_{x}+\sigma_{z})$ is the diagonal form of the tight-binding Hamiltonian, $\HTB$, from Eq.~\eqref{eqn:hamiltonian}, $\sigma_{z}$ is the $z$ Pauli matrix, $\vect{G}_{j}=\op{R}_{2j\pi/3}[0,\tfrac{2\pi}{\sqrt{3}a}]$, and $\gapmatrix$ describes the CDW gap. For simplicity, we take $\gapmatrix$ as constant and diagonal (no coupling between bands of different character). Guided by the effective description, Eq.~\eqref{eqn:effectiveh} in Appendix \ref{sec:LT}, we require $\gapmatrix_{11}<0$ and we find that we can obtain qualitative agreement with experiment by using a single parameter if $\gapmatrix=\gap\sigma_{z}$, $\gap<0$. In the theoretical plots in Fig.~\ref{fig:susceptibility} and Fig.~\ref{fig:comparison}, we have used $\gap=-60\,\mathrm{meV}$ and simulated the ARPES intensity using the wave vector-resolved spectral function,
% \begin{align*}
    %     A(\vect{k})=-\tfrac{1}{\pi}\sum_{\vect{k}}\Im\left[ \left(\HCDW-\sigma_{0}(\epsilon+i\eta)\right)^{-1} \right],
    %   \end{align*}
\begin{align}
  A(\vect{k})=-\tfrac{1}{\pi}\sum_{n}\Im\left[ \langle \vect{k},n \lvert \left(\HCDW-\epsilon-i\eta)\right)^{-1} \rvert \vect{k},n \rangle \right],
\end{align}
where $\Im\left[ x \right]$ stands for the imaginary part of $x$, $\ket{\vect{k},n}$ denotes the state with wave vector $\vect{k}$ in the band $n$ and $\eta$ is the phenomenological energy broadening. In Fig.~\ref{fig:susceptibility}(c), we have used $\eta=0.1\,\mathrm{eV}$ and in the bottom row of Fig.~\ref{fig:comparison}, $\eta=0.05\,\mathrm{eV}$.
    %     where $\sum_{\vect{k}}$ denotes a sum over the first two diagonal elements (corresponding to wave vector $\vect{k}$) and $\eta$ is the phenomenological energy broadening. In Fig.~\ref{fig:susceptibility}(c), we have used $\eta=0.1\,\mathrm{eV}$ and in the bottom row of Fig.~\ref{fig:comparison}, $\eta=0.05\,\mathrm{eV}$.

\section{Effective model of saddle point coupling}\label{sec:LT}

%%%%% Fig - effective model %%%%%
\begin{figure}[tb]
  \centering
  \includegraphics[width=1.00\columnwidth]{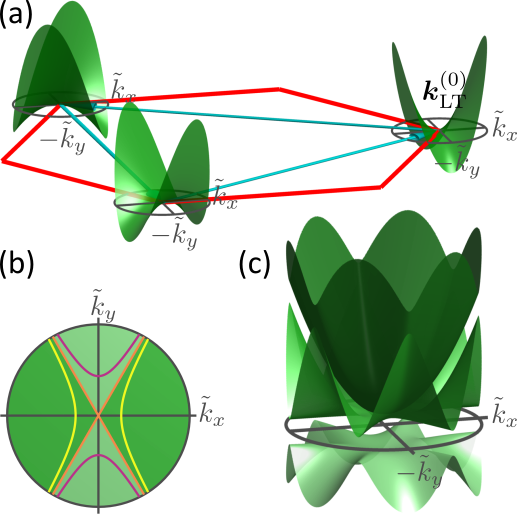}
  \caption{(a) Schematic of the minimal model for the coupling between the saddle points in the $\twobytwo$ charge density wave phase. The red hexagon represents the Brillouin zone of the $\twobytwo$ superlattice and the green surfaces show the electronic dispersion in the vicinity of three saddle points coupled by the superlattice reciprocal vectors indicated with the cyan arrows. (b) Constant-energy contours in the vicinity of a saddle point as described by $\op{H}_{\mathrm{LT}}(\tilde{\vect{k}})$ in Eq.~\eqref{eqn:hlteff} for the energies $\epsilon<\epsilon_{\mathrm{LT}}$ (yellow), $\epsilon=\epsilon_{\mathrm{LT}}$ (orange) and $\epsilon>\epsilon_{\mathrm{LT}}$ (purple). For $\epsilon=\epsilon_{\mathrm{LT}}$, one third of the electronic states are occupied as illustrated in light green. (c) The electronic dispersion in the vicinity of a saddle point reconstructed by attractive $\twobytwo$-periodic coupling.}
  \label{fig:effectivelt}
\end{figure}
%%%%% End of Fig - effective model %%%%%

Because of the energy separation between the bands, we only focus here on the band $\epsilon_{2}(\vect{k})$ which contains the saddle point of interest. For $\delta=0$, the dispersion in the vicinity of the saddle point at $\vect{k}^{(0)}_{\mathrm{LT}}$, shown in the right of Fig.~\ref{fig:effectivelt}(a), can be described by the effective Hamiltonian,
\begin{align}\label{eqn:hlteff}
  \hLT(\tilde{\vect{k}})=\frac{3a^{2}(t_{2}+\tilde{t}_{2})}{2}\left(3\tilde{k}_{x}^{2}-\tilde{k}_{y}^{2}\right),
\end{align}
where $\tilde{\vect{k}}=(\tilde{k}_{x},\tilde{k}_{y})=\vect{k}-\vect{k}^{(0)}_{\mathrm{LT}}$ is the wave vector measured from the saddle point (we discuss corrections to this minimal model arising from $\delta\neq 0$ in \cite{supplement}). As shown in light green in Fig.~\ref{fig:effectivelt}(b), for the Fermi level positioned exactly at such a saddle point, a third of the electronic states are occupied while two thirds remain empty. The $\twobytwo$ CDW-induced superlattice provides a new source of Bragg scattering which couples the saddle point at $\vect{k}^{(0)}_{\mathrm{LT}}$ to two other saddle points at $\vect{k}^{(2)}_{\mathrm{LT}}$ and $\vect{k}^{(4)}_{\mathrm{LT}}$. We present this coupling schematically in Fig.~\ref{fig:effectivelt}(a), with the red hexagon indicating the Brillouin zone of the $\twobytwo$ superlattice and the cyan arrows its basic reciprocal vectors which connect the saddle points. A simple description of this coupling is provided by the Hamiltonian,
\begin{align}\label{eqn:effectiveh}
\hLTCDW=\begin{bmatrix} \op{h}(\tilde{\vect{k}}) & \gaptwo & \gaptwo \\
\gaptwo & \op{h}(\op{R}_{\frac{2\pi}{3}}\tilde{\vect{k}}) & \gaptwo \\
\gaptwo & \gaptwo & \op{h}(\op{R}_{\frac{4\pi}{3}}\tilde{\vect{k}}) \end{bmatrix},
\end{align}
where the diagonal terms describe each of the three saddle points and $\gaptwo$ is the coupling due to the CDW order which we take to be real and a constant across the small area of the Brillouin zone that is of relevance. The Hamiltonian coupling saddle points at $\vect{k}^{(1)}_{\mathrm{LT}}$, $\vect{k}^{(3)}_{\mathrm{LT}}$ and $\vect{k}^{(5)}_{\mathrm{LT}}$ can be obtained by setting $\tilde{\vect{k}}\rightarrow-\tilde{\vect{k}}$ in $\op{H}_{\twobytwo}$. 

Diagonalizing $\hLTCDW$ exactly at the saddle point, $\tilde{\vect{k}}=0$, gives energies $\epsilon=2\gaptwo$ and $\epsilon=-\gaptwo$, with the latter double degenerate. For attractive interaction, $\gaptwo<0$, such a spectrum results in a gap at one third filling. The full spectrum of the Hamiltonian $\hLTCDW$ for $\gaptwo<0$ is presented in Fig.~\ref{fig:effectivelt}(c). Note that a gap separates the lowest band, shown in light green, from the higher ones, in agreement with experiment where no ARPES intensity is seen above the saddle point, Fig.~\ref{fig:arpes_maps}(j). As we demonstrate in the SM \cite{supplement}, a gap does not appear at one third filling in the case of the $\sqrtthree$ order.

\end{document}

%opening

\title{Supplemental Material: \\ Controlling charge density order in 2H-TaSe$_{2}$ using a van Hove singularity}
\author{W.~R.~B.~Luckin} % wl748@bath.ac.uk; will@luckin.co.uk
\thanks{These authors contributed equally to this work.}
\affiliation{Department of Physics, University of Bath, Claverton Down, Bath BA2 7AY, United Kingdom}
\author{Y.~Li} % liyw3@shanghaitech.edu.cn
\thanks{These authors contributed equally to this work.}
\affiliation{Institute for Advanced Studies (IAS), Wuhan University, Wuhan 430072, People’s Republic of China}
\affiliation{School of Physical Science and Technology, ShanghaiTech University, Shanghai 201210, People’s Republic of China}
\author{J.~Jiang} % jjiangcindy@ustc.edu.cn
%\affiliation{School of Physical Science and Technology, ShanghaiTech University, Shanghai 201210, People’s Republic of China}
%\affiliation{CAS-Shanghai Science Research Center, Shanghai 201210, People’s Republic of China}
%\affiliation{Hefei National Laboratory for Physical Sciences at the Microscale, University of Science \& Technology of China, Hefei 230026, People’s Republic of China}
\affiliation{School of Emerging Technology, University of Science and Technology of China, Hefei 230026, People’s Republic of China}
\author{S.~M.~Gunasekera} % surani04@hotmail.com
\affiliation{Department of Physics, University of Bath, Claverton Down, Bath BA2 7AY, United Kingdom}
\author{C.~Wen} % surani04@hotmail.com
\affiliation{School of Physical Science and Technology, ShanghaiTech University, Shanghai 201210, People’s Republic of China}
\author{Y.~Zhang} % surani04@hotmail.com
\affiliation{School of Physical Science and Technology, ShanghaiTech University, Shanghai 201210, People’s Republic of China}
\author{D.~Prabhakaran} % dharmalingam.prabhakaran@physics.ox.ac.uk
\affiliation{Department of Physics, University of Oxford, Oxford, OX1 3PU, United Kingdom}
\author{F.~Flicker} % flicker@cardiff.ac.uk
\affiliation{School of Physics and Astronomy, Cardiff University, Cardiff CF24 3AA, United Kingdom}
\author{Y.~Chen} % yulin.chen@physics.oxford.ac.uk; 
\affiliation{Department of Physics, University of Oxford, Oxford, OX1 3PU, United Kingdom}
\affiliation{School of Physical Science and Technology, ShanghaiTech University, Shanghai 201210, People’s Republic of China}
\affiliation{CAS-Shanghai Science Research Center, Shanghai 201210, People’s Republic of China}
\author{M.~Mucha-Kruczy\'{n}ski}
\email{M.Mucha-Kruczynski@bath.ac.uk}
\affiliation{Department of Physics, University of Bath, Claverton Down, Bath BA2 7AY, United Kingdom}
\affiliation{Centre for Nanoscience and Nanotechnology, University of Bath, Claverton Down, Bath BA2 7AY, United Kingdom}

\maketitle

\tableofcontents
\newpage

\beginsupplement

\section{Details of ARPES measurements}

	\subsection{General details and measurement geometry}

\added{For the data presented in the main text, the angle-resolved photoemission spectroscopy (ARPES) measurements on the pristine 2H-TaSe$_{2}$ were performed at beamline I05 (high-resolution ARPES branch) of Diamond Light Source (DLS), UK, while the ARPES measurements on the 2H-TaSe$_{2}$ with {\it{in situ}} potassium surface deposition were performed at beamline MERLIN of Advanced Light Source (ALS), USA. Similar experiment geometries were applied in measurements at both DLS and ALS, and linear horizontal (LH) polarization was used [the analyzer slit was vertical, as shown in Fig.~\ref{fig:arpes_geometry}(a); note our comment on polarization further in this section]. We note that we have measured the pristine samples both at DLS and ALS - we compare the corresponding Fermi surface maps in Fig.~\ref{fig:arpes_geometry}(b),(c). Both measurements show consistent results for pristine 2H-TaSe$_{2}$ samples, with the DLS map of slightly higher quality, and match with previous studies \cite{rossnagel_prb_2005, evtushinsky_prl_2008, borisenko_prl_2008}.

\begin{figure}[b]
\centering
\includegraphics[width=1.0\textwidth]{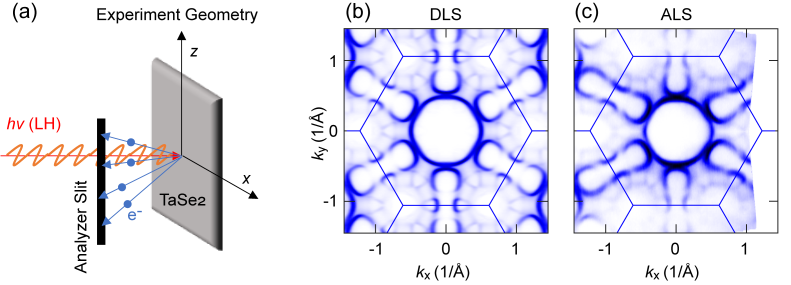}
\caption{\added{(a) Experimental geometry of measurements at DLS and ALS. The light polarization is linear horizontal and the analyzer slit is vertical ($z$-axis). For the measurement at DLS, the $\vect{\Gamma}$-$\vect{M}$ direction is aligned to the $z$-axis while the $\vect{\Gamma}$-$\vect{K}$ direction is aligned to the $x$-axis. For the measurement at ALS, the $\vect{\Gamma}$-$\vect{K}$ direction is aligned to the $z$-axis while the $\vect{\Gamma}$-$\vect{M}$ direction is aligned to the $x$-axis. (b),(c) Fermi surface mapping performed at DLS (b) and ALS (c).}}
\label{fig:arpes_geometry}
\end{figure}

In all ARPES measurements, the samples were cleaved {\it{in situ}} under ultra-high vacuum below $5\times 10^{-11}$ Torr and photons with energy of 80 eV were used. Data were collected by a Scienta R4000 analyzer at I05 beamline of DLS, UK, and a Scienta R8000 analyzer at beamline MERLIN of ALS, USA. The total energy resolution was better than 20 meV. The temperature of the sample during the measurements was kept at 130 K [Fig. 2(a) and (f)] and 10 K [Fig. 2(b) and (g)] for the high-temperature normal phase and CDW phase, respectively, at DLS. The temperature of the sample during the potassium surface deposition was kept at 22 K [Fig. 2(c-e) and (h-j)]. Alkali metal dispensers from SAES Getters were used for potassium evaporation which was conducted {\it{in situ}} with the sample at the measurement position. Evaporation current and time were precisely monitored ($5.2$ A and 60 s, 60 s and 120 s for the three doses between ARPES measurements).}

%\added{For the data presented in the main text, it is true that the measurements on the pristine samples were performed at DLS and the measurements on the K-doped ones were performed at ALS, although the pristine samples were measured at both DLS and ALS. Similar experiment geometries were applied in measurements from DLS and ALS, and linear-horizontal (LH) polarization was used (analyzer slit was vertical, as shown in Fig. R8(a)). Measurements from both DLS and ALS show consistent ARPES results on pristine 2H-TaSe2 samples, and matches with previous studies [Refs: Phys. Rev. Lett. 100, 236402 (2008); Phys. Rev. Lett. 100, 196402 (2008); Phys. Rev. B 72,121103 (2005)], as shown in Fig.~\ref{fig:arpes_geometry}.}

	\subsection{Data symmetrization}

\begin{figure}[b]
\centering
\includegraphics[width=1.0\textwidth]{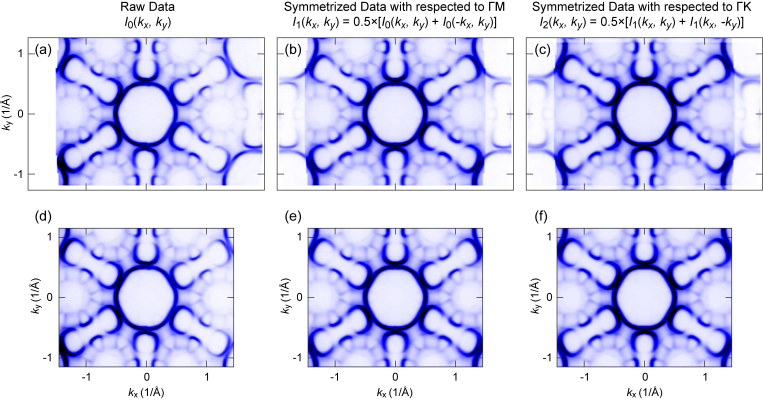}
\caption{\added{Symmetrization process. (a) Raw ARPES Data. (b) Data symmetrized with respect to $\vect{\Gamma}$-$\vect{M}$ ($k_{x}=0$). (c) Data symmetrized with respect to $\vect{\Gamma}$-$\vect{K}$ ($k_{y}=0$) following the symmetrization with respect to $k_{x}=0$. (d)-(f) Same as (a)-(c) but momentum ranges have been properly truncated.}}
\label{fig:arpes_symmetrization}
\end{figure}

\added{In Fig.~2 of the main text, we present ARPES maps which were symmetrized with respect to the mirror reflection lines along $\vect{\Gamma}$-$\vect{M}$ ($k_{x}=0$) and $\vect{\Gamma}$-$\vect{K}$ ($k_{y}=0$), based on the crystalline symmetry of the material. We present this procedure and its impact on the raw data in Fig.~\ref{fig:arpes_symmetrization}.} 

	\subsection{Polarization dependence of the ARPES intensity}

\begin{figure}[b]
\centering
\includegraphics[width=1.0\textwidth]{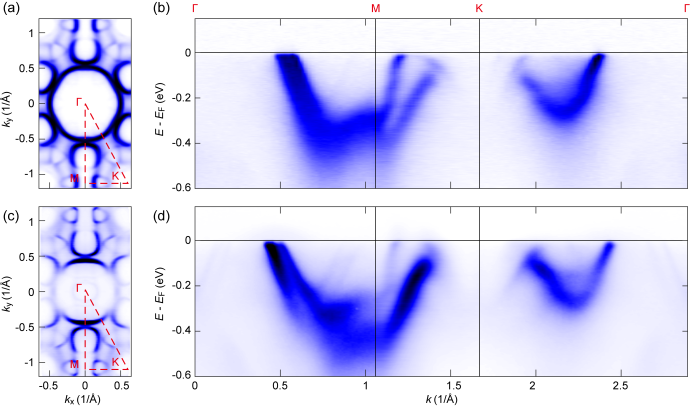}
\caption{\added{Polarization-dependent ARPES measurements. (a) and (c), Fermi surface contours. (b) and (d), band dispersion along the $\vect{\Gamma}$-$\vect{M}$-$\vect{K}$-$\vect{\Gamma}$ directions. (a) and (b) are measured using photons with linear horizontal polarization, energy $\hbar\omega=80$ eV. (c) and (d) are measured using photons with linear vertical polarization, $\hbar\omega=50$ eV.}}
\label{fig:arpes_polarization}
\end{figure}

\added{As stated earlier in this section, the data presented in the main text was obtained using linear horizontal (LH) polarization of the incoming light. We have measured ARPES intensities using both LH and linear vertical (LV) polarized light. In Fig.~\ref{fig:arpes_polarization}, we compare the Fermi surface contours and band dispersions along the high-symmetry directions measured for the undoped sample in the $\threebythree$ CDW using both of the polarizations. One can see that, although the spectral weight distributions measured with these two polarizations are different, the measured band structures are consistent. We have checked that interpretation of the measured photoemission intensity is consistent between the two light polarizations. This also gives us confidence that modulations of intensity we interpret as opening of new gaps due to CDW do not originate instead from polarization or matrix element effects.}

\section{Potassium deposition}

\begin{figure}[b]
\centering
\includegraphics[width=0.7\textwidth]{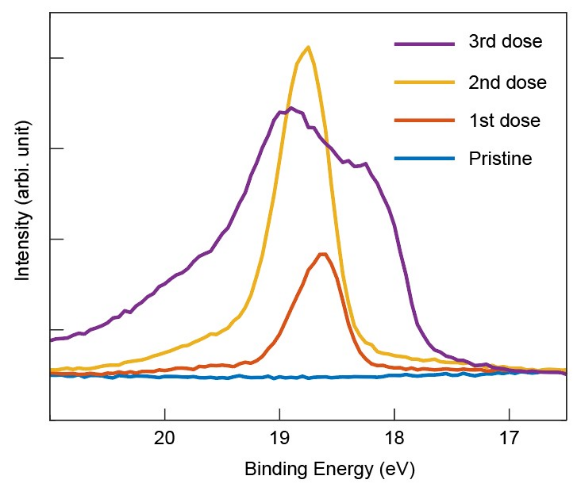}
\caption{Evolution of the core level spectranear the potassium core level peak upon surface potassium doping.}
\label{fig:potassium}
\end{figure}

\begin{figure}
\centering
\includegraphics[width=1.0\textwidth]{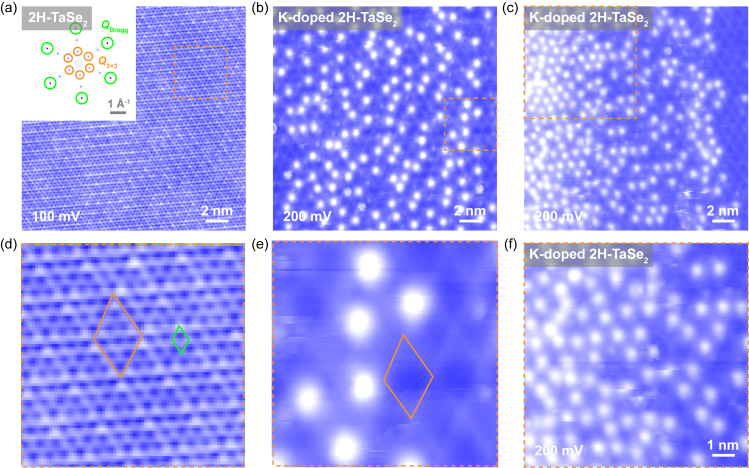}
\caption{\added{STM measurements on (a) the pristine 2H-TaSe$_{2}$, {\it{in-situ}} potassium-doped surface of 2H-TaSe$_{2}$ with (b) low dopant concentration and (c) high dopant concentration. The inset in (a) shows the corresponding fast Fourier transform pattern. (d)-(f) Zoom-in plot of areas in (a)-(c), as indicated by the orange squares. The green rhombus and circles denote the normal state unit cell and Fourier peaks, respectively. The pink rhombi and circles denote the $\threebythree$ supercell and Fourier peaks, respectively.}}
\label{fig:stm}
\end{figure}

\added{
Fig.~\ref{fig:potassium} shows the evolution of core level spectra near the potassium core level peak as a function of potassium coverage following its depositions. One can see that the pristine surface is free of potassium atoms, whereas after the first and second doses of potassium the spectra show a single-peak feature. This indicates that the coverage is less than or comparable to one layer. In contrast, the spectra exhibit a clear double-peak profile after the third dose, suggestive of the formation of a second layer of potassium on the surface. We note that the new CDW order emerges after the last, third dose of potassium, when the atoms have formed more than just one layer on the surface of TaSe$_{2}$. This suggests that the potential ordering of potassium on the surface is not driving the spectral changes we have observed. This is consistent with our scanning transmission microscopy studies of the surface following potassium doping, see Sec.~\ref{sec:stm} and Fig.~\ref{fig:stm}, in which we do not observe any long-range ordering of the deposited atoms. 
} 
 
\added{
To mention, using our tight-binding model, we estimate the doping necessary to shift the Fermi level from the Fermi level of the pristine material to the saddle point as $n\sim 0.19$ electron per unit cell, or $\sim 0.1$ electron per TaSe$_{2}$ formula unit (this assumes a rigid band model in the normal state and does not account for any band renormalization). This is a significant additional charge density in the system and its presence may be contributing to the destruction of the $\threebythree$ order. At the same time, as we are working with a bulk crystal, the $\threebythree$ order should persist far enough from the surface where the impact of deposited potassium can be neglected. To mention, the generalized electronic susceptibility we show in Fig.~4 of the main text shows that the single-particle dispersion features a prominent peak indicating the $\threebythree$ instability, pretty much independently of the Fermi level. This is consistent with the fact that NbSe$_{2}$, which has qualitatively a similar band structure but the Fermi level above the saddle point (not because of the additional charge but rather due to changes in the shape of the bands), also displays the (incommensurate) $\threebythree$ CDW.
}

\added{
\section{Scanning transmission microscopy studies of the surface of the sample}\label{sec:stm}

We have carried out scanning transmission microscopy (STM) measurements on the potassium-doped surface of 2H-TaSe$_{2}$, as shown in Fig.~\ref{fig:stm}. One can see that the pristine 2H-TaSe$_{2}$, panels (a) and (b), shows a very nice $\threebythree$ CDW pattern. The $\threebythree$ order persists at low potassium dosing level, panels (b) and (e). In the area with larger amount of potassium atoms, panels (c) and (f), we are unable to resolve the charge order below. According to our estimation of the doping level based on the core-level spectra as shown in Fig.~\ref{fig:potassium}, the new correlated phase emerges when potassium atoms start to form the second layer on the surface, which makes it impossible to detect using STM.
}

\added{
\section{LEED measurements}

Apart from the scanning transmission microscopy, we also used low-energy electron diffraction (LEED) to study our surface. LEED uses diffraction of a collimated beam of low-energy electrons to provide information about the surface of the sample. We point out that the charge density order is weak as compared to the original lattice order. As a consequence, very few CDW orders are visible in LEED. To the best of our knowledge, no LEED peaks of the (3×3) CDW order have been reported for TaSe$_{2}$ despite this CDW having been thoroughly investigated for decades. Our proposed (2×2) CDW is to be expected to be even more difficult to detect than the (3×3) order as the expected signal will be affected by the disordered potassium on the surface.

\begin{figure}
\centering
\includegraphics[width=1.0\textwidth]{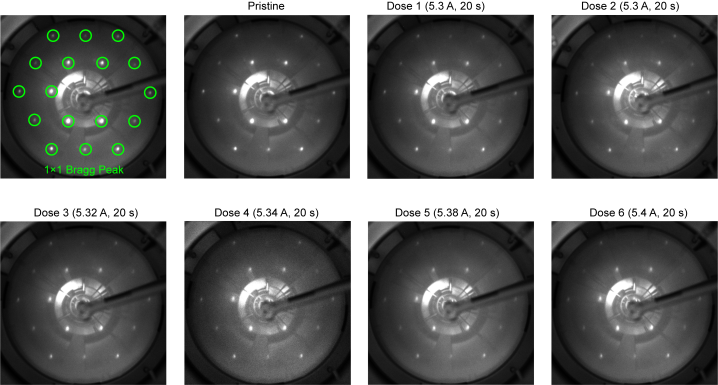}
\caption{\added{LEED measurements of the pristine 2H-TaSe$_{2}$ surface as well as following six consecutive {\it{in situ}} surface potassium dosing processes.}}
\label{fig:leed}
\end{figure}

In Fig.~\ref{fig:leed}, we show LEED measurements of 2H-TaSe$_{2}$ single crystals with pristine surface as well as following six consecutive {\it{in situ}} potassium surface doping stages. As expected, only $1\times1$ Bragg peaks of the original lattice have been observed and they become blurred upon the surface doping process, due to the disordered introduced by potassium on the surface.
}

\section{Importance of out-of-plane dispersion}

\begin{figure}
\centering
\includegraphics[width=0.8\textwidth]{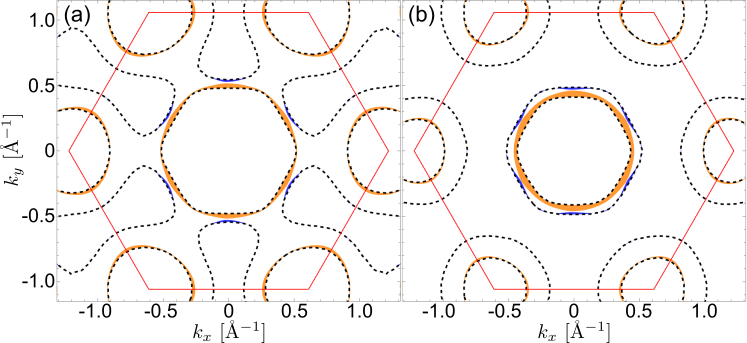}
\caption{Projections of, (a), the three-dimensional Fermi surface, in panel (b), constant-energy surface for $\epsilon=\epsilon_{0}+0.15$ eV, of 2H-TaSe$_{2}$ on the two-dimensional Brillouin zone. The boundary of the Brillouin zone is shown in solid red, the dashed black lines indicate the bands at $k_{z}=0$ and the orange and blue areas show the parts of the band structure with the largest $k_{z}$ dispersion for each of the two bands.}
\label{fig:kz}
\end{figure}

In Fig.~\ref{fig:kz}, we show the projections of, in panel (a), the three-dimensional Fermi surface, in panel (b), constant-energy surface for $\epsilon=\epsilon_{0}+0.15$ eV, on the two-dimensional Brillouin zone. Such projections demonstrate the small impact of out-of-plane dispersion on the overall band structure which is limited to slight broadening of the contours discussed in the main text. Importantly, the change of topology occurs as a function of energy as described in the main manuscript for any choice of $k_{z}$. This justifies our decision to limit the theoretical model to a planar cut for a constant $k_{z}$.

\begin{figure}[t]
\centering
\includegraphics[width=1.0\textwidth]{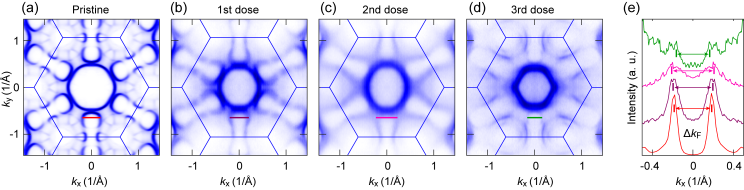}
\caption{\added{Comparison of the width of the dogbone electron pocket at the Fermi level, $\Delta k_{\mathrm{F}}$, upon surface potassium doping. (a)-(d) ARPES maps from panels (b)-(e) of Fig.~2 of the main text. (e) A comparison of the wave vectors $\Delta k_{\mathrm{F}}$ of the dog-bone electron pockets at each doping level, as indicated by the solid lines in (a)-(d).}}
\label{fig:spectral_added}
\end{figure}

\added{
\section{Confirming the occurence of Lifshitz transition under potassium deposition}

As discussed in the main text, in the normal state, 2H-TaSe$_{2}$ has two band sheets in the vicinity of the Fermi level. In the presence of the charge density order (of any periodicity), appearance of a larger supercell associated with this order intuitively leads to folding of the normal state bands into a smaller Brillouin zone and hence a larger number of bands is expected. This leads to novel features in photoemission intensity and complicates the analysis of band dispersion. Here, we comment further on some of the features of the spectra presented in Fig.~2 of the main text. %In panel (e) of Fig.~2 of the main text, there is some spectral intensity connecting the contours surrounding $\vect{K}$ and $\vect{\Gamma}$. This intensity can also be seen in panel (j), above and to the left and right of the bright flat intensity feature in the top band. We argue here that these features are due to the charge density order and consistent with our scenario of $\twobytwo$ periodicity.
} 
\addedd{In Fig.~\ref{fig:spectral_added}(a)-(d), we have reproduced the panels (b)-(e) from Fig.~2 of the main text. In (d), weak intensity connecting the $\vect{\Gamma}$ and $\vect{K}$ pockets can be noticed. Similarly, in the top row of Fig.~\ref{fig:spectral_added_2}, we reproduce panels (g)-(j) of Fig.~2 of the main text. In the right-most panel corresponding to the sample after the third deposition of potassium, weak intensity features are seen at wave vectors $\pm 0.2$ \AA$^{-1}$ at energies above the bright intensity spot indicating the top of the high-energy band. This weak intensity corresponds to the same features connecting the $\vect{\Gamma}$ and $\vect{K}$ pockets. We argue here that these features are due to the charge density order and consistent with our scenario of $\twobytwo$ periodicity.
}

\added{We point out that, in general, since the CDW super-potential is much weaker than the original periodic lattice potential, it follows that spectral intensity of CDW-folded bands should be much weaker than the original “main” bands. This is, for example, the case for the weak triangular features in Fig.~2(b) of the main text [reproduced as panel (a) in Fig.~\ref{fig:spectral_added}] also highlighted with arrows in Fig.~3(a) of the main text, a well-known signature of the $\threebythree$ CDW \cite{li_prb_2018}. Similarly, the M-shape bands in Fig.~2(g) of the main text [reproduced as the left-most panel of the top row of Fig.~\ref{fig:spectral_added_2}] are also due to the $\threebythree$ CDW \cite{li_prb_2018}. We note that no reconstruction is expected around the saddle points in the $\threebythree$ CDW because of the mismatch between the CDW wave vector and the relative positions of the saddle points [note that in Fig.~2(b) of the main text, also Fig.~\ref{fig:spectral_added}(a), reconstruction due to the $\threebythree$ CDW induces minigaps at the centre of the dogbones while minimally affecting the dogbone edges responsible for the Lifshitz transition].

\begin{figure}[t]
\centering
\includegraphics[width=0.9\textwidth]{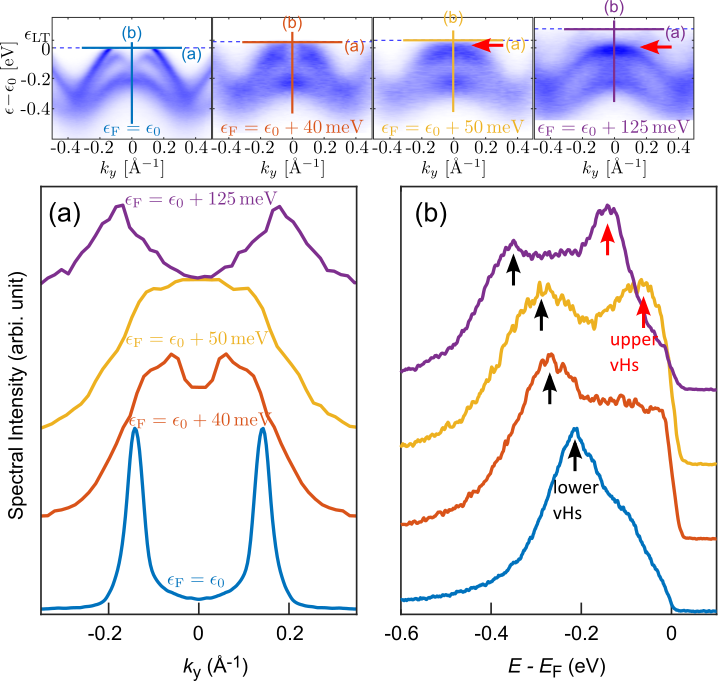}
\caption{\addedd{Top row: same as panels (g)-(j) of Fig.~2 of the main text. (a) Momentum distribution curves (MDCs) at the Fermi energy $\epsilon_{\mathrm{F}}$ for the three different potassium doping levels along the momentum direction as shown in the top panel. (b) Energy distribution curves (EDCs) for the three different potassium doping levels at the position of the saddle point, $(k_{x}=-\tfrac{2\pi}{3a}-\tfrac{4}{\sqrt{5}a}\delta,k_{y}=0)$.}}
\label{fig:spectral_added_2}
\end{figure}

Given the topology of the electronic bands in the high-temperature normal state as shown in Fig.~1 of the main text, one can easily track the ``main'' Fermi surface and band evolution in between the surface potassium depositions. As the Fermi level increases, the $\vect{\Gamma}$- and $\vect{K}$-centered hole pockets must shrink and the $\vect{M}$-centered dogbone electron pockets must expand, until the Fermi level crosses the saddle point. If $\epsilon_{\mathrm{F}}$ of Fig.~2(e) and (j) of the main text [panel (d) of Fig.~\ref{fig:spectral_added} and the right-most panel in the top row of Fig.~\ref{fig:spectral_added_2}, respectively] is still below the saddle point and the weak, string-like intensity feature connecting the $\vect{\Gamma}$ and $\vect{K}$ pockets is part of the dogbone electron pocket, then this implies that the dogbone electron pocket in Fig.~2(e) is bigger than that in Fig.~2(b)-(d). On the contrary, as illustrated in Fig.~\ref{fig:spectral_added}, in which in panel (e) we compare the Fermi momenta $\Delta k_{\mathrm{F}}$ of the dogbone electron pockets at each doping level, the Fermi momentum $\Delta k_{\mathrm{F}}$ of the assumed dogbone electron-pocket in panel (d) [Fig.~2(e) of the main text] is clearly smaller than that in panels (a)-(c) [Fig.~2(b)-(d) of the main text].
}

\begin{figure}[t]
\centering
\includegraphics[width=0.75\textwidth]{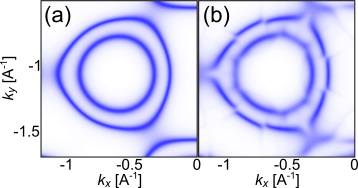}
\caption{\addedd{Simulated intensity maps at the energy $\epsilon=\epsilon_{0}+125$ meV in the vicinity of the $\vect{K}$ point for the high-temperature normal state (left) and the $\twobytwo$ CDW (right).}}
\label{fig:spectral_added_3}
\end{figure}

\addedd{
To add to this argument, in Fig.~\ref{fig:spectral_added_2}, we show, in panel (a), the momentum distribution curves (MDCs) at the Fermi level for the three different potassium dopings along the momentum direction perpendicular to $\vect{\Gamma}$--$\vect{K}$ and passing through the saddle point [the same direction as the cuts reproduced in the top row of Fig.~\ref{fig:spectral_added_2} from Fig.~2(g)-(j) of the main text]. One can unambiguously see that the neighboring dogbone shoulders move closer as the Fermi energy changes from $\epsilon_{\mathrm{F}}=\epsilon_{0}$ (blue) to $\epsilon_{\mathrm{F}}=\epsilon_{0}+40$ meV (red) and $\epsilon_{\mathrm{F}}=\epsilon_{0}+50$ meV (yellow), indicative of approaching the vHs. For $\epsilon_{\mathrm{F}}=\epsilon_{0}+125$ meV (purple), we observe peak features which correspond to the additional spectral features connecting the $\vect{\Gamma}$ and $\vect{K}$ pockets. However, since the momentum separation between the two peaks is significantly larger than that measured for lower Fermi levels, these spectral features cannot be attributed to the “main bands” unless opposite experimental evidence is given. Attributing them to the “main bands” and concluding that the Fermi level is still below the saddle point leads to inconsistencies when looking at other spectral features. To demonstrate this, we show in panel (b) of Fig.~\ref{fig:spectral_added_2} the energy distribution curves (EDCs) for the three different potassium doping levels at the position of the saddle point, $(k_{x}=-\tfrac{2\pi}{3a}-\tfrac{4}{\sqrt{5}a}\delta,k_{y}=0)$. For the last two dopings, $\epsilon_{\mathrm{F}}=\epsilon_{0}+50$ meV and $\epsilon_{\mathrm{F}}=\epsilon_{0}+125$ meV (yellow and purple, respectively), a new intensity feature is visible close to the Fermi level, marked with the red arrow. This feature corresponds to the fact that the bright intensity spot marked with the red arrow in the two right-most panels of the top row of Fig.~\ref{fig:spectral_added_2} extends across $k_{y}=0$ and in our interpretation can be easily understood as movement of the saddle point of the high-energy ``main band” below the Fermi level. This feature cannot be easily reconciled, if at all, with the idea that the saddle point is not yet below the Fermi level because the weak features connecting the $\vect{\Gamma}$ and $\vect{K}$ pockets originate from the main bands rather than are due to the CDW super-potential.
}

\addedd{
Finally, in Fig.~\ref{fig:spectral_added_3}, we compare the simulated intensity maps at the energy $\epsilon=\epsilon_{0}+125$ meV in the vicinity of the $\vect{K}$ point for the high-temperature normal state (left) and the $\twobytwo$ CDW (right). The latter corresponds to the map shown in Fig.~4(c) of the main text but here, we decreased the broadening to $\eta=50$ meV (we used the same broadening in the left panel). For this Fermi level, the saddle point is occupied in both cases. For the normal state, this is clear from the two contours enclosing the $\vect{K}$ point. For the case of the $\twobytwo$ CDW, the decrease of the broadening shows more detailed features of the intensity profile. Importantly, the outside contour displays weak intensity features that ``connect'' it to the contour around $\vect{\Gamma}$. This is fully consistent with our experimental observations and confirms that i) our scenario of the Fermi level above the saddle point after the last potassium doping and formation of the $\twobytwo$ CDW is in agreement with all observed intensity features and that ii) the weak intensity features connecting the $\vect{\Gamma}$ and $\vect{K}$ pockets and observed for the sample with the highest potassium doping are due to the $\twobytwo$ CDW super-potential.
}

\added{
\section{Symmetry of $\threebythree$-induced features around $\vect{K}$}

\begin{figure}[b]
\centering
\includegraphics[width=1.0\textwidth]{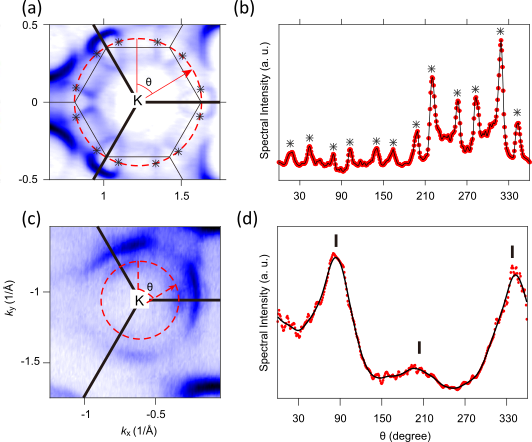}
\caption{\added{C$_{6}$ and C$_{3}$ symmetry of the electronic structure modifications due to the $\threebythree$ and $\twobytwo$ orders around $\vect{K}$. (a)  Raw ARPES data for the Fermi surface of the pristine TaSe$_{2}$ in the $\threebythree$ CDW state in the vicinity of $\vect{K}$. (b) Spectral intensity distribution around $\vect{K}$, as indicated by the red dashed circle in (a). Corresponding peaks are marked by asterisks in (a) and (b). Panels (d) and (e) are similar to (b) and (c), respectively, but for the $\twobytwo$ phase.}}
\label{fig:symmetry}
\end{figure}

The spectral intensity of ARPES features is closely related to the magnitude of the corresponding periodic order \cite{voit_science_2000} as well as the complicated matrix element effect. For TaSe$_{2}$ the dominant periodic potential leads to C$_{3}$ symmetry around $\vect{K}$ and so one naturally expects to observe such symmetry in the vicinity of the Brillouin zone corner in the normal state, as can be seen in Fig.~2(a) of the main text. Moreover, one also expects vestiges of the C$_{3}$-symmetric main bands in the CDW phase -- these are clearly visible in the outer band in Fig.~3(a) of the main text. Here, we additionally show the as measured ARPES data in the vicinity of $\vect{K}$ in Fig.~\ref{fig:symmetry}(a). 
}

\added{
Ultimately, when looking for ARPES features due to (some) CDW order, we are interested in modifications of the normal state electronic band structure. The electronic structure of the $\threebythree$ order has been well studied, both in terms of the CDW-folded Fermi surface \cite{li_prb_2018} as well as gap distributions \cite{rossnagel_prb_2005}, and our observations are in full agreement with other experiments. This includes the appearance of weak C$_{6}$ symmetry in the vicinity of $\vect{K}$ (see in particular Fig.~2(e) in \cite{rossnagel_prb_2005}). To further support this point, we plot in Fig.~\ref{fig:symmetry}(b) the angular dependence of the spectral intensity along the red dashed circle in Fig.~\ref{fig:symmetry}(a). The twelve peaks correspond to the triangular features in panel (a), indicating weak C$_{6}$ symmetry of the new, CDW-folded parts of the Fermi surface around $\vect{K}$. In contrast to the $\threebythree$ state, the modification of the electronic structures in the new correlated state shows no signs of similar, weak C$_{6}$ symmetry. We demonstrate this in panels (c) and (d) of Fig.~\ref{fig:symmetry}, which show the raw ARPES data in the vicinity of $\vect{K}$ as well as the angular dependence of the spectral intensity for the inner band. In general, the new order gaps the double-walled $\vect{K}$ pockets into 3 sections and the gap structure clearly displays C$_{3}$-symmetric features only, as we have elaborated in Fig.~3(b) and Fig.~5 of the main text.
}

\section{Spectral weights in the vicinity of $\vect{K}$ for $\threebythree$ and $\sqrtthree$ phases}

\begin{figure}[b]
\centering
\includegraphics[width=1.0\textwidth]{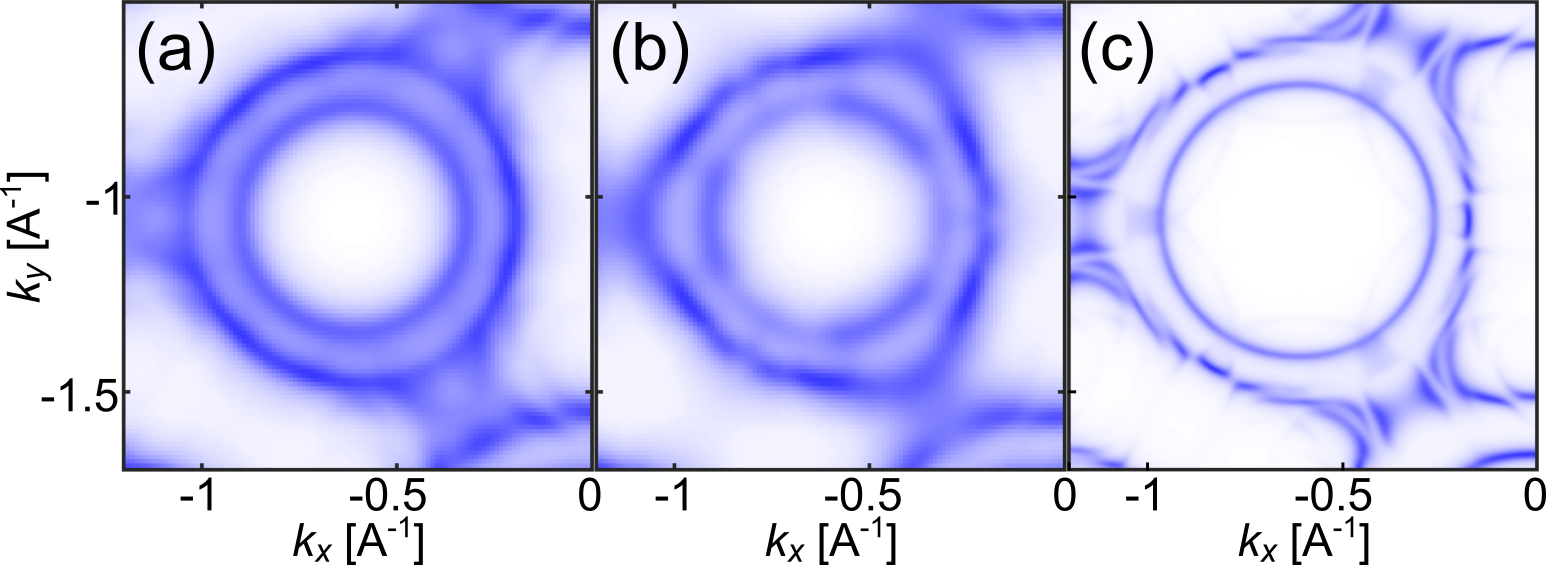}
\caption{(a), (b) Spectral weights in the vicinity of the $\vect{K}$ Brillouin zone corner computed at the energy $\epsilon_{0}+0.125$ eV for the (a) $\threebythree$ and (b) $\sqrtthree$ superlattice models. \added{(c) Spectral weight in the vicinity of $\vect{K}$ for the $\threebythree$ superlattice at the energy $\epsilon_{0}$.}}
\label{fig:spectral}
\end{figure}

In Fig.~4 of the main text, we showed the spectral function in the vicinity of $\vect{K}$ computed at the energy $\epsilon_{0}+0.125$ eV for the $\twobytwo$ CDW. In Fig.~\ref{fig:spectral}(a) and (b), we show the equivalent figures for the $\threebythree$ and $\sqrtthree$ superlattices. They have been modelled similarly to the $\twobytwo$ phase, the model for which is described in the main text and its appendices, by considering Bragg scattering by the new assumed periodicity. We have only assumed scattering between bands of the same character and used a single, real constant to describe the magnitude of the coupling. While such an approach is very simple, a comparison between Fig.~\ref{fig:spectral} and Figs. 3(b) and 4(c) of the main text demonstrates that the $\threebythree$ and $\sqrtthree$ phases are fundamentally different from the experimental constant-energy map. The $\threebythree$ phase does not display the right number of arcs in either of the rings around $\vect{K}$. The $\sqrtthree$ phase shows six intensity maxima in the outer ring while the positions of the three inner ring maxima are reversed with respect to experimental observation. In both cases, the inconsistencies are underpinned by the drastically different way of folding the single-particle bands as compared to the $\twobytwo$ phase rather than details of the model. \added{Note that panel (a) of Fig.~\ref{fig:spectral} should not be directly compared to Fig.~3(a) of the main text as the latter is for the Fermi level $\epsilon_{\mathrm{F}}=\epsilon_{0}=\epsilon_{\mathrm{LT}}-50$ meV whereas in panel (a) we simulate how the $\threebythree$ state might appear in ARPES if it existed in the potassium doped sample with $\epsilon_{\mathrm{F}}=\epsilon_{0}+125\,\mathrm{meV}=\epsilon_{\mathrm{LT}}+75$ meV. For completeness, we show in Fig.~\ref{fig:spectral}(c) simulated ARPES intensity for the $\threebythree$ superlattice at the energy $\epsilon_{\mathrm{F}}=\epsilon_{0}$. We have used small broadening to show that our simple, single-parameter model does present the most important features of the $\threebythree$ state: gaps along the longer edge of the dogbone contours as well as weakly six-fold-symmetric features around the Brillouin zone corner. The multitude of folded bands generates some additional features that are not present or are strongly washed out in the experiment; moreover, the intensity of the inner pocket is, in our model, only weakly affected by the new super-periodicity. However, there is an overall qualitative agreement with the experiment.}  

\section{Saddle point - effective model}

As mentioned in the main manuscript, the saddle points of interest to this work are located at
\begin{align}
\vect{k}^{(n)}_{\mathrm{LT}}=\op{R}_{n\pi/3}\vect{k}^{(0)}_{\mathrm{LT}}=\op{R}_{n\pi/3}\left(\frac{2\pi}{3a}+\frac{4}{a\sqrt{5}}\delta\right)\begin{bmatrix} 1 \\ 0 \end{bmatrix},
\end{align}
where $\op{R}_{\theta}$ is the operator of rotation by angle $\theta$, $a$ is the lattice constant, $n=0,\ldots,5$ and
\begin{align}
\delta=\frac{t_{1}+\tilde{t}_{1}}{6(t_{2}+\tilde{t}_{2})}.
\end{align}
The energy at the saddle point is
\begin{align}
\epsilon_{\mathrm{LT}} = (t_{0}+\tilde{t}_{0})-2(t_{2}+\tilde{t}_{2})(1-3\delta+12\delta^{2}-4\delta^{3}),
\end{align}
which in the main manuscript has been set to zero by our choice of the tight-binding parameters. To compare, at the $\vect{M}$ point,
\begin{align}
\epsilon_{\vect{M}}=(t_{0}+\tilde{t}_{0})-2(t_{2}+\tilde{t}_{2})(1+6\delta),
\end{align}
so that the saddle points at $\vect{k}^{(0)}_{\mathrm{LT}}$ and $\vect{M}$ are shifted with respect to each other by
\begin{align}
\Delta\epsilon=2(t_{2}+\tilde{t}_{2})(9\delta-12\delta^{2}+4\delta^{3})=(t_{1}+\tilde{t}_{1})\left(3-4\delta+\frac{4}{3}\delta^{2}\right).
\end{align}
Notice, that for $\delta=0$ and hence, $t_{1}+\tilde{t}_{1}=0$, $\Delta\epsilon=0$ so that the saddle points at $\vect{k}^{(n)}_{\mathrm{LT}}$ and $\vect{M}$ are at the same energy. This might make it difficult to distinguish between the influence of the two as the Fermi level is tuned past $\epsilon_{\mathrm{LT}}=\epsilon_{\vect{M}}$, especially as reciprocal vectors of a $\twobytwo$ charge density wave nests not only the saddle points at $\vect{k}^{(n)}_{\mathrm{LT}}$ but also those at $\vect{M}$ onto each other.

In order to describe the nesting of the saddle points at $\vect{k}^{(n)}_{\mathrm{LT}}$ in the $\twobytwo$ and $\sqrtthree$ superlattices, we expand the tight-binding Hamiltonian in the vicinity of the point $\vect{Q}=\tfrac{2\pi}{3a}(1,0)$,
\begin{align}\label{eqn:heff}
\op{h}(\tilde{\vect{k}})=\alpha \left(\tilde{k}_{x}-\delta\right)^{2}-\beta \tilde{k}_{y}^{2},
\end{align}
where $\tilde{\vect{k}}=(\tilde{k}_{x},\tilde{k}_{y})$ is the wave vector measured from $\vect{Q}$ and the coefficients $\alpha$ and $\beta$ are,
\begin{align}\begin{split}\label{eqn:coefficients}
\alpha & = \frac{3a^{2}(t_{2}+\tilde{t}_{2})}{2}\left(3+\delta-8\delta^{2}+4\delta^{3}\right), \\
\beta & = \frac{3a^{2}(t_{2}+\tilde{t}_{2})}{2}\left(1+3\delta-4\delta^{3}\right),
\end{split}\end{align}
which gives
\begin{align}
\frac{\alpha}{\beta}\approx3-8\delta+16\delta^{2}.%\rightarrow\sqrt{\left\vert\frac{\alpha}{\beta}\right\vert}\approx\sqrt{3}\left(1-\frac{4}{3}\delta+\frac{22}{9}\delta^{2}\right).
\end{align}
For $\delta=0$, our expressions take the form as provided in the main text and the point of expansion becomes the position of the saddle point, $\vect{Q}=\vect{k}^{(0)}_{\mathrm{LT}}$. The dispersion in the vicinity of the saddle points at $\vect{k}^{(n)}_{\mathrm{LT}}$, $n=1,\dots,5$, can be obtained by using the fact that the electronic dispersion of 2H-TaSe$_{2}$ has $C_{6}$ rotational symmetry.

\section{Comparison of $\twobytwo$ and $\sqrtthree$ phases}

For both the $\twobytwo$ and $\sqrtthree$ superlattices, the saddle points at $\vect{k}^{(n)}_{\mathrm{LT}}$ are nested onto each other, albeit not ideally for $\delta\neq 0$. We recall that, as shown in Fig.~6 of the main text, for $\delta=0$ and the Fermi level at the saddle point, within the effective description of Eq.~\eqref{eqn:heff} [and the coefficients as in Eq.~\eqref{eqn:coefficients}], a third of states is occupied and two thirds are empty. For $\delta\neq 0$, this relationship is only very weakly affected. Assuming that the impact of the superlattice is dominant in the vicinity of the saddle points and the rest of the 2H-TaSe$_{2}$ dispersion is only weakly modified, this implies that after nesting 3 saddle points for the $\twobytwo$ phase, the position of the Fermi level is such that one full band of the effective model is occupied. In turn, for the $\sqrtthree$ phase in which all 6 saddle points nest onto each other, two bands should be fully occupied. In this Section, we show that at these Fermi levels, a gap can be realised at the saddle points for the $\twobytwo$ superlattice, in agreement with the experimental observations (Fig.~2 and 5 of the main text). However, this is not the case for the $\sqrtthree$ phase. 

%The argument presented here for the \((2 \times 2)\) CDW rests on the existance of an observed gap in the spectrum at \(\vect{k}_{\mathrm{LT}}\) in the doped ARPES measurements (Figure 2, main text). For the simple quadratic expansion in the previous Section, placing the saddle point exactly at zero energy, it can be seen that for \(\alpha = -3\beta\) the number of states below zero energy divided by the number above gives the filling fraction \(1/3\), meaning that for composite systems consisting of replicas of this dispersion the Fermi level will also be found where \(1/3\) of the states are occupied.

%Using this argument, we show that for the $(2\times 2)$ CDW it is possible to observe a gap at the Fermi energy, and for the $(2\sqrt{3}\times 2\sqrt{3})$ CDW it is not.

\subsection{$\twobytwo$ CDW}

\begin{figure}
  \centering
  \includegraphics[width=0.7\textwidth]{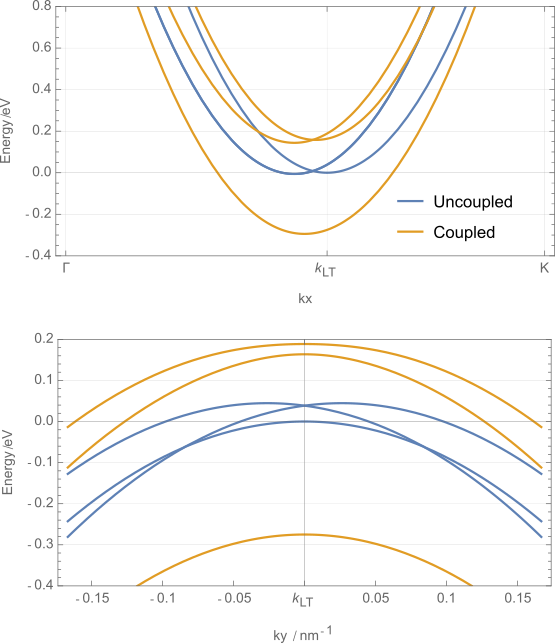}
  \caption{The band structure of the effective model of three saddle points at $(\vect{k}^{(0)}_{\mathrm{LT}}, \vect{k}^{(2)}_{\mathrm{LT}}, \vect{k}^{(4)}_{\mathrm{LT}})$ coupled by superlattice vector $\vect{q}_{\twobytwo}=\vect{M}$ [a $\twobytwo$ CDW], along the $\tilde{k}_{x}$ (top) and $\tilde{k}_{y}$ (bottom) axes.}
  \label{fig:Mspec}
\end{figure}

The $\twobytwo$ CDW introduces a superlattice vector \(\vect{Q}=\vect{M}\), which couples the Lifshitz points at \(\vect{k}_{n}\) in two inequivalent sets of three, $(\vect{k}^{(0)}_{\mathrm{LT}}, \vect{k}^{(2)}_{\mathrm{LT}}, \vect{k}^{(4)}_{\mathrm{LT}})$ and $(\vect{k}^{(1)}_{\mathrm{LT}}, \vect{k}^{(3)}_{\mathrm{LT}}, \vect{k}^{(5)}_{\mathrm{LT}})$. Considering the first set and defining \(\deltak = \frac{4}{a\sqrt{5}}\delta\), we can write the Hamiltonian for these three coupled points evaluated exactly at \(\vect{k}^{(0)}_{\mathrm{LT}}\) as 
\begin{equation}
  \op{h}_{\twobytwo}(\vect{k}^{(0)}_{\mathrm{LT}}) = 
  \begin{pmatrix}
    0 & \Delta & \Delta \\
    \Delta^* & \frac{9 \alpha  \deltak^2}{4}-\frac{3 \beta \deltak^2}{4} & \Delta \\
    \Delta^* & \Delta^* & \frac{9 \alpha \deltak^2}{4}-\frac{3 \beta \deltak^2}{4} \\
  \end{pmatrix},
\end{equation}
with some effective coupling \(\Delta\) due to the CDW superlattice. For real $\Delta$, eigensolving yields energies 
\begin{equation}
  \epsilon^{\twobytwo}_1 = 2\Delta + \frac{1}{2} (3 \alpha - \beta) \deltak^2,
\end{equation}
\begin{equation}
  \epsilon^{\twobytwo}_2 = -\Delta + \frac{1}{4} (3 \alpha - \beta) \deltak^2,
\end{equation}
\begin{equation}
  \epsilon^{\twobytwo}_3 = -\Delta + \frac{3}{4} (3 \alpha - \beta) \deltak^2,
\end{equation}
taking corrections only to \(O(\deltak^2)\). One can see that for $\Delta<0$, occuping \(1/3\) of states whilst leaving a gap at the Fermi energy is possible, as \(\epsilon_1\) is well seperated in energy from \(\epsilon_2\) and \(\epsilon_3\). To confirm that, we solve the system numerically away from \(\vect{k}_0\) and show the resulting band structure along the $\tilde{k}_{x}$ and $\tilde{k}_{y}$ axes in the top and bottom of Figure~\ref{fig:Mspec}, respectively (in the figure, we have used $\Delta=-0.15$ eV).

\subsection{$\sqrtthree$ CDW}

The $\sqrtthree$ CDW introduces a superlattice vector \(\vect{q}_{\sqrtthree}=1/2\vect{K}\), which couples the Lifshitz transition points at \(\vect{k}^{(n)}_{\mathrm{LT}}\) in a ring of six. The Hamiltonian for this system at $\vect{k}^{(0)}_{\mathrm{LT}}$ can be written
\begin{equation}
  \label{eq:lt_6x6h}
  \op{h}_{\sqrtthree}(\vect{k}^{(0)}_{\mathrm{LT}}) =
  \begin{pmatrix}
     0 & \Delta & 0 & 0 & 0 & \Delta \\
 \Delta^* & \frac{\alpha \deltak^2}{4}-\frac{3 \beta \deltak^2}{4} & \Delta & 0 & 0 & 0 \\
 0 & \Delta^* & \frac{9 \alpha \deltak^2}{4}-\frac{3 \beta \deltak^2}{4} & \Delta & 0 & 0 \\
 0 & 0 & \Delta^* & 4 \alpha \deltak^2 & \Delta & 0 \\
 0 & 0 & 0 & \Delta^* & \frac{9 \alpha \deltak^2}{4}-\frac{3 \beta \deltak^2}{4} & \Delta \\
 \Delta^* & 0 & 0 & 0 & \Delta^* & \frac{\alpha \deltak^2}{4}-\frac{3\beta \deltak^2}{4} \\
  \end{pmatrix}.
\end{equation}
\begin{figure}
  \centering
  \includegraphics[width=0.7\textwidth]{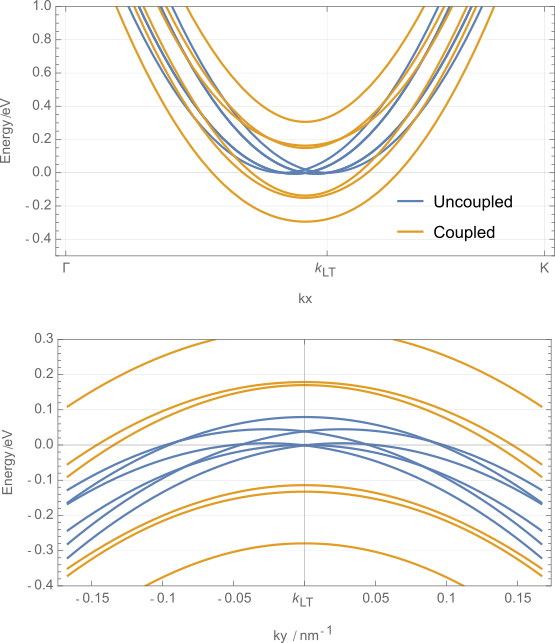}
  \caption{The bandstructure of the effective model of six saddle points at $\vect{k}^{(n)}_{\mathrm{LT}}$, $n=0,1,\dots,5$, coupled by superlattice vector $\vect{q}_{\sqrtthree}=1/2\vect{K}$ [a $\sqrtthree$ CDW], for cuts along along the $\tilde{k}_{x}$ (top) and $\tilde{k}_{y}$ (bottom) axes.}
  \label{fig:Kspec}
\end{figure}
Again, assuming real $\Delta$, this system has eigenenergies
\begin{equation}
  \epsilon^{\sqrtthree}_1 = -2\Delta + \frac{1}{2} (3 \alpha - \beta) \deltak^2,
\end{equation}
\begin{equation}
  \epsilon^{\sqrtthree}_2 = -\Delta + \frac{1}{4} (5 \alpha - 3 \beta) \deltak^2,
\end{equation}
\begin{equation}
  \epsilon^{\sqrtthree}_3 = -\Delta + \frac{1}{4} (7 \alpha - \beta) \deltak^2,
\end{equation}
\begin{equation}
  \epsilon^{\sqrtthree}_4 = \Delta + \frac{1}{4} (5 \alpha - 3 \beta) \deltak^2,
\end{equation}
\begin{equation}
  \epsilon^{\sqrtthree}_5 = \Delta + \frac{1}{4} (7 \alpha - \beta) \deltak^2,
\end{equation}
\begin{equation}
  \epsilon^{\sqrtthree}_6 = 2\Delta + \frac{1}{2} (3 \alpha - \beta) \deltak^2,
\end{equation}
with accuracy up to \(O(\deltak^2)\). This spectrum is fundamentally different to that of the $\twobytwo$ CDW in that no gap exists at \(1/3\) filling due to the closeness of $\epsilon^{\sqrtthree}_2$ and $\epsilon^{\sqrtthree}_3$ (for $\Delta>0$) or $\epsilon^{\sqrtthree}_4$ and $\epsilon^{\sqrtthree}_5$ (for $\Delta<0$). Again, numerically solving away from \(\vect{k}^{(0)}_{\mathrm{LT}}\) confirms these arguments hold in the proximal region, as can be seen in Figure~\ref{fig:Kspec} (in Fig.~\ref{fig:Kspec}, we have used $\Delta=-0.15$ eV).

\section{Electronic density of states in $\twobytwo$ phase}

\begin{figure}
  \centering
  \includegraphics[width=0.7\textwidth]{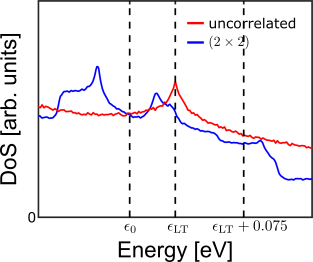}
  \caption{Electronic density of states (DoS) of the uncorrelated normal state (red) and the proposed $\twobytwo$ charge density wave (blue) as produced using the Hamiltonians described in the main text. The two curves have been aligned with respect to each other so that for the Fermi level at the energy $\epsilon=\epsilon_{\mathrm{LT}}$, the number of states occupied is the same for both phases.}
  \label{fig:dos}
\end{figure}

In Fig.~\ref{fig:dos}, we compare the electronic density of states of the predicted $\twobytwo$ CDW, $\rho_{\twobytwo}(\epsilon)$, and the uncorrelated normal state, $\rho(\epsilon)$, in the vicinity of the saddle point, $\epsilon=\epsilon_{\mathrm{LT}}$. As anticipated, the van Hove singularity present at $\epsilon=\epsilon_{\mathrm{LT}}$ in the uncorrelated phase is pushed to lower energies in the $\twobytwo$ phase which lowers the overall density of states at the Fermi level. We have also computed the total energy of the electrons in the two bands contributing to the Fermi surface and have confirmed that the total energy decreases in the $\twobytwo$ phase as compared to the uncorrelated state -- that is,
\begin{align}
\int_{-\infty}^{\tilde{\epsilon}}\epsilon\rho_{\twobytwo}(\epsilon)d\epsilon<\int_{-\infty}^{\epsilon_{\mathrm{LT}}}\epsilon\rho(\epsilon)d\epsilon,
\end{align}
with the energy $\tilde{\epsilon}$ given by the condition of the same number of states occupied in both phases,
\begin{align}
\int_{-\infty}^{\tilde{\epsilon}}\rho_{\twobytwo}(\epsilon)d\epsilon = \int_{-\infty}^{\epsilon_{\mathrm{LT}}}\rho(\epsilon)d\epsilon.
\end{align}

\added{
While our discussion of the electronic densities of states agrees with our qualitative arguments, we note here some limitations of our approach. We are describing the CDW phases using a simple and, ultimately, single-particle model based on Bragg scattering on a new superlattice. While this allows a basic description of the electronic dispersion in the vicinity of the Fermi level, it i) relies on a single parameter describing the CDW gap across the whole Brillouin zone and at all energies, ii) ignores the energetics of the associated lattice distortion. In the case of the latter point, it is clear that the lattice should, in principle, be included in a proper comparison of stabilities of competing correlated orders. The former, in turn, is relevant because, generically, in the case of degenerate perturbation theory with a constant coupling, electronic states at the edges of the spectrum are pushed away the most (to also note, construction of an effective model means we lack states below the two bands we describe which could push these bands upwards). This impacts the calculation of total energy which needs to include states from the bottom of the band. For these reasons, we limited ourselves to a brief comparison of the $\twobytwo$ order and the normal state only, with the main being the qualitative lowering of the electronic density of states within the right energy range. A meaningful comparison of the correlated orders with each other is a much more complicated question to answer. 
}